\documentclass[usenatbib]{mnras}

\usepackage{etoolbox}
\newtoggle{postreferee}
\togglefalse{postreferee}

\newtoggle{postrefereeB}

\newtoggle{devmode}

\newtoggle{removesomefits}

\newtoggle{CompareCentres}

\usepackage{footmisc}

\usepackage{newtxtext,newtxmath}

\usepackage[T1]{fontenc}
\usepackage{ae,aecompl}

\newcommand{\projecttitle}{Detecting cosmic voids via maps of geometric-optics parameters}
\newcommand{\projectversion}{e4f7af0}
\newcommand{\projectzenodoid}{zenodo.7792910}
\newcommand{\projectzenodohref}{\href{https://zenodo.org/record/7792910}{zenodo.7792910}}
\newcommand{\projectzenodohrefShowURL}{\href{https://zenodo.org/record/7792910}{https://zenodo.org/record/7792910}}
\newcommand{\projectzenodofilesbase}{https://zenodo.org/record/7792910/files}
\newcommand{\projectgitrepository}{\url{https://codeberg.org/mpeper/lensing}}

\newcommand{\projectgitrepositoryarchived}{\href{https://archive.softwareheritage.org/swh:1:rev:b5dff23ab8ba8c758112d5fd3f737fb6f44cd6fe\%3Borigin=https://codeberg.org/mpeper/lensing}{swh:1:rev:b5dff23ab8ba8c758112d5fd3f737fb6f44cd6fe}}

\newcommand{\revolvername}{{\sc revolver}}

\newcommand{\OmegaMvalue}{0.3}

\newcommand{\OmegaLvalue}{0.7}

\newcommand{\Hubblevalue}{70.0}

\newcommand{\Ncrootvalue}{256}

\newcommand{\Lboxvalue}{120}

\newcommand{\rockstarparamMINHALOOUTPUTSIZEvalue}{10}

\newcommand{\sageparamPartMassvalue}{0.86}

\newcommand{\MainVoidRedshiftvalue}{0.5}

\newcommand{\ObserverVoidDistancevalue}{1322.0}

\newcommand{\ToleranceFracMeanvalue}{0.90}

\newcommand{\VoidStddevFracCritvalue}{0.50}

\newcommand{\VoidStddevFracCritThetavalue}{0.30}

\newcommand{\VoidStddevFracCritSigmavalue}{0.30}

\newcommand{\VoidDiffFracCritvalue}{1.50}

\newcommand{\VoidDiffFracCritThetavalue}{1.30}

\newcommand{\VoidDiffFracCritSigmavalue}{1.10}

\newcommand{\StdRVoidProbvalue}{0.30}

\newcommand{\MergeDistancevalue}{10.0}

\newcommand{\StdXzVoidProbvalue}{5.0}

\newcommand{\CosmicRayPercentilevalue}{90}

\newcommand{\NPixelsvalue}{120}

\newcommand{\NAveragevalue}{5}

\newcommand{\NNeighboursvalue}{5}

\newcommand{\MassDefNVoidvalue}{28}

\newcommand{\MassDefNVoidGammavalue}{30}

\newcommand{\MassDefNVoidRevolvervalue}{46}

\newcommand{\MassDefNAveragePixelvalue}{25}

\newcommand{\OSNVoidExpvalue}{34}

\newcommand{\OSNVoidSigmavalue}{39}

 \newcommand\VoidMatchSignificanceTable{
    \Sigma   & 0.017   & 0.00051   & 0.94   & 0.94 \\
 \bargamma   & 0.00010   & 1.0\times 10^{-5}   & 0.93   & 0.53 \\
    \theta   & 0.00022   & 1.0\times 10^{-5}   & 0.48   & 0.16 \\
\vert\sigma\vert   & 7.0\times 10^{-5}   & 1.0\times 10^{-5}   & 0.43   & 0.37 \\
}

\newcommand{\XzMedianGivenThreeDUniqMassDefvalue}{7.8}

\newcommand{\XzMedianGivenThreeDUniqGammavalue}{6.8}

\newcommand{\XzMedianGivenThreeDUniqExpvalue}{5.8}

\newcommand{\XzMedianGivenThreeDUniqSigvalue}{4.5}
  \newcommand{\machinearchitecture}{x86\_64}
\newcommand{\machinebyteorder}{Little Endian}

 \togglefalse{postrefereeB}   \togglefalse{removesomefits}   \togglefalse{devmode}   \toggletrue{CompareCentres}

\usepackage{amsmath}

\usepackage{csquotes}

\iftoggle{postreferee}{

  \usepackage{color}  \definecolor{myred}{rgb}{0.7,0.0,0.2}
}{

}

\iftoggle{postrefereeB}{

  \usepackage{color}  \definecolor{myred}{rgb}{0.7,0.0,0.2}
}{

}

\usepackage{graphicx}

\usepackage[font={footnotesize}]{caption}

\usepackage{xcolor}

\usepackage{tipa}

\newcommand{\mktab}[1]{\textcolor{black!30!white}{\_\_\_TAB\_\_\_}}

\providecommand\prd{Physical Review D}

\providecommand\physrep{Phys.Rep.}

\providecommand\aap{A\&A}

\providecommand\apj{ApJ}
\providecommand\apjs{ApJS}
\providecommand\mnras{MNRAS}
\providecommand\apjl{ApJL}
\providecommand\apjl{Astrophys.J.Lett.}

\providecommand\SPIEConfSeries{Society of Photo-Optical Instrumentation Engineers (SPIE) Conference Series}
\providecommand\SPIEConfSeries{SPIE Conf. Ser. }

\providecommand\CiSE{Comp.~in~Sci.~Eng.}

\providecommand\SSS{Sect.~}
\providecommand\SSSs{Sections~}
\providecommand\diffd{\mathrm{d}}

\newcommand\gtapprox{\,\lower.6ex\hbox{$\buildrel >\over \sim$}\,}
\newcommand\ltapprox{\,\lower.6ex\hbox{$\buildrel <\over \sim$}\,}

\newcommand{\rmpi}{\mathrm{\pi}}

\newcommand\Ommzero{\Omega_{\mathrm{m0}}}   \newcommand\OmLamzero{\Omega_{\Lambda0}}

\newcommand{\tildeSigma}{\widetilde{\Sigma}}

\newcommand{\barX}{\overline{X}}
\newcommand{\barSigma}{\overline{\Sigma}}
\newcommand{\bargamma}{\overline{{\gamma}_{\perp}}}

\newcommand{\doublemeanX}{\overline{\overline{X}}}

\newcommand{\doublemeangamma}{\overline{\overline{{\gamma}_{\perp}}}}

\newcommand{\meanabsslopeX}{\overline{\left\lvert{\overline{X}'}\right\rvert}}

\newcommand{\thrD}{three-dimensional} \newcommand{\twD}{two-dimensional}

\DeclareMathOperator\erf{erf}
 \iftoggle{devmode}{
}{
}

\title[Cosmic voids from geometric optics]{\projecttitle}

\author[Peper, Roukema \& Bolejko]{Marius Peper,$^{1}$
         Boudewijn F. Roukema,$^{1,2}$
         Krzysztof Bolejko$^{3}$
         \\
         $^1$ Institute of Astronomy, Faculty of Physics,
         Astronomy and Informatics, Nicolaus Copernicus
         University, Grudziadzka 5, 87-100 Toru\'n, Poland\\
         $^2$ Univ Lyon, Ens de Lyon, Univ Lyon1, CNRS, Centre de
         Recherche Astrophysique de Lyon UMR5574, F--69007, Lyon,
         France\\
         $^3$ School of Natural Sciences, College of Sciences and Engineering,
         University of Tasmania, Private Bag 37 Hobart TAS 7001, Australia}

\date{Accepted \ldots Received \ldots; in original form \ldots}
\pubyear{2023}

\begin{document}

\maketitle

\begin{abstract}
{Curved-spacetime geometric-optics maps derived from a deep photometric survey should contain information about the three-dimensional matter distribution and thus about cosmic voids in the survey, despite projection effects.}
{We explore to what degree sky-plane geometric-optics maps can reveal the presence of intrinsic {\thrD} voids.}
{We carry out a cosmological $N$-body simulation and place it further than a gigaparsec from the observer, at redshift $\MainVoidRedshiftvalue$.
    We infer {\thrD} void structures using the watershed algorithm.
    Independently, we calculate a surface overdensity map and maps of weak gravitational lensing and geometric-optics scalars.
    We propose and implement a heuristic algorithm for detecting (projected) radial void profiles from these maps.}
{We find in our simulation that given the sky-plane centres of the {\thrD} watershed-detected voids, there is significant evidence of correlated void centres in
    the surface overdensity $\Sigma$,
    the averaged weak-lensing tangential shear $\bargamma$,
    the Sachs expansion $\theta$, and
    the Sachs shear modulus $\lvert\sigma\rvert$.
    Recovering the centres of the {\thrD} voids from the sky-plane information alone is significant given the weak-lensing shear $\bargamma$, the Sachs expansion $\theta$, or the Sachs shear $\lvert\sigma\rvert$, but not significant for the surface overdensity $\Sigma$.
    Void radii are uncorrelated between {\thrD} and {\twD} voids; our algorithm is not designed to distinguish voids that are nearly concentric in projection.}
{This investigation shows preliminary evidence encouraging observational studies of gravitational lensing through individual voids, either blind or with spectroscopic/photometric redshifts.
    The former case -- blind searches -- should generate falsifiable predictions of intrinsic {\thrD} void centres.}
\end{abstract}


\section{Introduction}
Spectroscopic redshifts to determine the (comoving) three-dimensional structures of cosmic voids require much more telescope resources than photometric surveys alone.
While the simplest interpretation of a (single-filter) photometric survey is that it shows only the projected galaxy positions and shapes on the sky, the fact that the Universe is inhomogeneous implies that the photometric map contains information on the {\thrD} distribution of inhomogeneities, with effects that are generically referred to as gravitational lensing.
Gravitational lensing theory was developed many decades ago in its generic form of geometric optics \citep{Sachs1961,Sasaki1993}, and gravitational lensing by overdensities was detected with the twin quasar QSO 0957+561 A/B in 1979 \citep{WalshCarswellW1979,YoungGunn1980,Gorenstein1984,Gorenstein1988}, and with giant luminous arcs \citep{Paczynski1987} and an Einstein cross \citep{AdamBacon1989EinCross}, providing visually striking evidence favouring general relativity.
For an in-depth review, see \citet{BartelmannSchneider01}.
Here, we argue that geometric-optics parameters that are derived from a deep photometric extragalactic map should contain information that can be used to detect some of the physical, {\thrD} cosmic voids in the map, despite the fact that the voids are projected on the sky plane together with foreground and background voids.
By carrying out a full $N$-body simulation and analysing it, using an {\it a priori} reproducible software method, we aim to explore to what degree maps of photometrically derived geometric-optics parameters can reveal intrinsic {\thrD} voids.
We consider both the conventional approximation of weak-lensing parameters and Sachs optical scalars derived directly from the evolving gravitational potential.

The discovery of comsic voids goes well back over 40 years \citep{GregoryThompson78, Joeveer1978VoidDiscovery}, from galaxy surveys that indicated that large regions appear to be devoid from galaxies, with galaxies being located primarily in structures that are usually described as walls, filaments and clusters.
Modern observations indicate that the comoving volume of our Universe is dominated by cosmic voids.
For example, measurements based on the Sloan Digital Sky Survey (SDSS) indicate that a total fraction of roughly $60\%$ of the volume consists of cosmic voids \citep{Pan2012voids}.
Cosmic voids have recently gained in interest, as they provide different characteristics for testing cosmological models to those provided by overdense structures.
\cite{Baojiu2012fRvoids} use simulations to study voids in modified gravity models, namely a fifth force that would affect the size and the density of voids.
Dark energy is suspected to influence the shape of a void \citep{Bos2012DEVoids}.
Studying the shape, density profile, size and abundance of cosmic voids should yield crucial information about our Universe \citep[e.g.][]{Peebles01Voids,NadHot2013,Pisani15voidcounts}.

Multiple strategies for detecting voids in either observations or simulations exist.
Since voids consist of the absence of luminous matter rather than its presence, this is a challenging task.
Several different methods are commonly used for tracing these underdense regions of the cosmic web.
Early routines used the assumption of spherical structures \citep[e.g.][]{Kauffmann91VoidFinder} to detect voids.
This was justified from the theoretical description of a void evolving out of a tophat-filtered density fluctuation; ellipsoidal initial density profiles were found to generally evolve to become more spherical \citep{GunnGottComa72, Lilje1991Voids, Sheth04}.
The watershed void finders \citep[e.g.][]{Neyrinck2008Zobov, AragonCalvo10CosmicSpine} make no assumptions on the shape of a void and is close to being parameter-free.
This has become a {\it de facto} standard for determining the shapes of cosmic voids.
Watershed mechanisms detect local minima in the density distribution of the cosmic web and identify underdense structures by searching for successively higher density contours, effectively finding the overdense edges of voids.
The properties of the resulting voids depend on the spatial number density of the tracer particles used to represent the matter distribution.
\citet{NadathurHotch15voidsI} show that randomly subsampling the density of dark matter particles will tend to bias the void statistics, and suggest the use of halo occupation distribution models instead.
To apply a watershed void finder to observational galaxy data, spectroscopic redshifts are needed.

\cite{SanchezTwoDVoidFinding} introduce a method of detecting voids from a multi-filter photometric survey by analysing redshift slices whose thickness is based on the photometric redshift uncertainties.
The first detector variable that we analyse here results in a roughly comparable method.
We detect structures in the surface overdensity, which, under the approximation of a constant mass-to-luminosity ratio, can be inferred from the observed photometric survey.

However, we are primarily interested in other sky-plane variables that can yield information on the large-scale structure of the Universe: the gravitional lensing signal.
Due to the nature of dark matter having very weak interactions apart from gravity, we cannot measure the dark matter distribution of the Universe directly from electromagnetic surveys.
Photometric galaxy surveys are generally thought to provide a fair proxy for the real projected matter distribution, but with many caveats.
In contrast, the gravitational interaction of the dark matter distribution with photons can be measured via lensing effects.
A light bundle that transverses a cosmological structure will experience shear and expansion and in particular, cosmic voids should leave an imprint on the shear and expansion.
This should make it possible to reconstruct the underdensity field of voids based on the lensing signal, without no dependence on assumptions about baryon cooling or star formation.
Large-scale maps of the gravitational lensing signal could thus, in principle, be used to detect cosmic voids.
This would provide a method independent of using the projected spatial distribution of galaxies, since the lensing signal depends on the full underlying mass density, no matter whether it is luminous or not.

While in this work we consider lensing parameters that are measurable from source galaxies far beyond the voids that we aim to detect, other observational methods of constructing the lensing signal have been proposed.
\cite{Lewis06CMBLensing} argue that since the cosmic microwave background (CMB) is lensed, the lensing potential can be reconstructed based on the observed CMB power spectrum.
Another method was proposed by \cite{Croft18LyAlphaLensing} to use Lyman~$\alpha$ forest observations to obtain lensing signals in the foreground of redshift slices of the forest.

In this work, we present a software pipeline (which aims to be fully reproducible on any unix-like operating system with sufficient RAM and disk space; \citealt{Akhlaghi2020maneage}) to generate a cosmological $N$-body simulation, to detect galaxies and voids in it, and to ray-trace geometrics-optics parameters.
The source package is provided as a frozen record at {\projectzenodoid}\footnote{\label{footnote-zenodo}{\projectzenodohrefShowURL}} and in live\footnote{\label{footnote-codeberg}\projectgitrepository} and archived\footnote{\projectgitrepositoryarchived} {\sc git} repositories.

In \SSS\ref{s-method-pipeline} we briefly describe our overall pipeline, extending that used in \citet{PeperRoukema2020}.
We describe our simulation geometry in \SSS\ref{s-method-geometry}.
We first detect intrinsic {\thrD} voids using the watershed algorithm (\SSS\ref{s-method-3D-voids}).
We independently try to detect voids in projection, \enquote*{photometrically} (in the absence of spectroscopic and photometric redshift information), from either the surface overdensity, conventional weak-lensing or other geometric-optics signals.
Our generic void profile search algorithm is defined in \SSS\ref{s-method-photo-voids}.
We compare the sky-plane positions and radii of the photometric voids to those of the intrinsic {\thrD} voids, using Monte Carlo simulations to check if this association is better than random (\SSS\ref{s-method-matches}).
In \SSS\ref{s-model} we present our four detector variables, including a modification of our default void profile search algorithm specific to the weak-lensing shear, in \SSS\ref{s-meth-lensing-gamma}).
We present our results in \SSS\ref{s-results}, discuss these in \SSS\ref{s-discuss} and conclude in \SSS\ref{s-conclu}.
This version of the paper was produced with git commit {\projectversion} of the source, after downloading, configuring, compiling and running on a computer with a {\machinebyteorder} {\machinearchitecture} architecture.

\section{Method}

\subsection{Software pipeline} \label{s-method-pipeline}

We use a highly reproducible software pipeline, following the Maneage template for reproducibility \citep{Akhlaghi2020maneage}, that generates a realistic distribution of galaxies using a succession of several different cosmological tools.
The software pipeline extends the galaxy formation pipeline presented in \citet{PeperRoukema2020}.
The pipeline includes a full simulation chain, starting with the generation of initial conditions with {\sc mpgrafic} and running an $N$-body simulation with {\sc RAMSES} \citep{PrunetPichon08mpgrafic, Teyssier02}. This simulation is processed as in \citet{1993ASPC...51..298R,RPQR97}, but using more recent software packages: dark matter haloes are detected and their merger-history tree is built with {\sc Rockstar} and {\sc consistent-trees} \citep{Behroozi13rockstar, Behroozi13MHT} and semi-analytical galaxy formation recipes are applied with {\sc SAGE} \citep{Croton16SAGE}.
We detect voids traced by the resulting spatial distribution of galaxies using a watershed void finder with {\revolvername} \citep{Nadathur19BOSSRevolver}.
The coding of our sky-plane void-profile search algorithm (\SSSs\ref{s-method-photo-voids} and~\SSS\ref{s-meth-lensing-gamma}) is original to this paper.
The full details of the analysis are provided in a live git repository\footref{footnote-codeberg} and a frozen Zenodo record\footref{footnote-zenodo}.
We used fixed pseudo-random seeds in most steps in the pipeline, as indicated in the configuration files.
A detailed discussion of the pipeline can be found in \citet{PeperRoukema2020}; here we primarily focus on new steps that are added to the analysis.

\subsection{Simulation geometry} \label{s-method-geometry}

We use a simulation for a standard $\Lambda$CDM model (cold dark matter cosmological model with a cosmological constant $\Lambda$) with Hubble constant $H_0 = {\Hubblevalue}$, current dark energy parameter $\OmLamzero = {\OmegaLvalue}$ and current matter density $\Ommzero= {\OmegaMvalue}$, $N_{\mathrm{part}}= {\Ncrootvalue}^3$ particles and a comoving box size $L_{\mathrm{box}} = {\Lboxvalue}\, \mathrm{Mpc}/h$, where $h := H_0/(100 \mathrm{km/s/Mpc})$.
This yields a dark matter particle mass of $m_{\mathrm{DM}} = {\sageparamPartMassvalue} \times 10^{10} ~ M_{\odot}/h$.
We require at least ${\rockstarparamMINHALOOUTPUTSIZEvalue}$ particles to detect a halo.
Since our simulated volume is a standard 3-torus ($T^3$) simulation, we detect dark matter haloes, generate a merger-history tree and detect intrinsic voids by interpreting the simulation's spatial section as $T^3$.

For computational convenience, for the \enquote*{observational} steps in which we detect voids in a sky-plane map of a detector variable, we interpret the (projected or ray-traced) simulated volume as the fundamental domain of a 2-torus ($T^2 := S^1 \times S^1 \times \mathbb{R}$), where the two multiply connected directions lie in the sky plane.
We informally refer to the fundamental domain as the \enquote*{box}.
The foreground and background of the box, at lower and higher redshifts, respectively, are implicitly assumed to be a homogeneous (structure-free), simply connected $\Lambda$CDM background, i.e., they are assumed to be transparent and flat, with no effect on gravitational lensing.
This simplification helps focus on the primary questions of our analysis; future analyses should include the effects of the full past time cone.
We assume that this box is observed at high redshift, $z' = \MainVoidRedshiftvalue$, corresponding to a large distance from the observer compared to the box size, $\chi_{\mathrm{O}} := {\ObserverVoidDistancevalue}$~Mpc$/h$ in comoving units.
We use both an observer-centred Euclidean comoving coordinate system, with the observer at $(x,y,z)=(0,0,0)$ and the centre of the simulated volume on the $y$ axis of the $\Lambda$CDM simply connected space at $(0,\chi_{\mathrm{O}},0)$, and a simulation-centred system shifted by $\chi_{\mathrm{O}}$.
To model light rays detected by the observer we convert $(x,y,z)$ to
$( x = \chi \sin\theta' \cos\varphi',
  y = \chi \sin\theta' \sin\varphi',
  z = \chi \cos\theta')$,
where $\chi$, the comoving radial distance, together with $\theta' \in [0,\rmpi]$ and $\varphi'$ are spherical coordinates of the spatial part of a flat $\Lambda$CDM model, and the simulation's centre is at $(\chi= \chi_{\mathrm{O}}, \,\theta'=\rmpi/2, \,\varphi'=\rmpi/2)$.
We compute each of our detector variables on a grid with $N_\mathrm{grid}={\NPixelsvalue}^2$ \enquote*{pixels} that we place on the middle plane of the box ($y={\ObserverVoidDistancevalue}$~Mpc$/h$), at $(x,z)$ positions in the grid.
We model light rays emitted from an observer-centred spherical surface near the back face of the simulated box ($T^2$ slice), through to a second observer-centred spherical surface close to the front face of the box.
The light rays's spatial paths are assumed to be straight in the non-perturbed space, i.e. with constant $\theta'$ and $\varphi'$.
We avoid approximately 5\% of the front and back parts of the box to minimise edge effects.
Projected variables for a given pixel are computed along the line of sight of a light ray passing from the back spherical surface, through the pixel, to the front spherical surface.
In other words, the grid approximately corresponds to what is often referred to as \enquote*{the sky plane} for small solid angles, although it is (in this construction) a genuinely flat plane in comoving space.
For brevity, we will use the term \enquote*{sky plane} as equivalent to this grid.

For the optical scalar calculations (\SSS\ref{s-meth-opt-scalars}), we trace our light rays geometrically under the assumption of a flat $\Lambda$CDM model, but calculate the optical scalars with a longitudinal Newtonian gauge approximation of an inhomogeneous model, with the line element
\begin{equation}
  \mathrm{d}s^2 = a^2 \left[-(1 + 2\Phi) \mathrm{d}\tau^2 + (1 - 2\Phi) \left(\mathrm{d}\chi^2 + \chi^2 \mathrm{d}\Omega^2  \right)  \right] \,,
\end{equation}
where $\tau$ is conformal time, $\Phi$ is a potential, and $\diffd \Omega$ is the solid spherical angle element $(\diffd\theta')^2 + (\cos\theta' \diffd\phi')^2$.

\subsection{Void detection} \label{s-method-voids}
\subsubsection{Intrinsic {\thrD} voids} \label{s-method-3D-voids}
We detect intrinsic voids traced by the galaxy population using the void finder {\revolvername}, which is based on the watershed void finder {\sc zobov} \citep{Neyrinck2008Zobov,Nadathur19BOSSRevolver}.
The watershed mechanism in {\sc zobov} uses a Voronoi tessellation to estimate the densities at the particles' positions, is nearly parameter-free and makes no assumptions on the shape of the void.
To characterise the size of a void, we use the effective radius $R_{\mathrm{eff}}$, which is based on the sum over the volumes $V_i$ of all the Voronoi cells that constitute a void, i.e. $R_{\mathrm{eff}} := \frac{3}{4 \rmpi} \left( \sum_i V_i \right)^{1/3}$.
We adopt the geometric centroid of the set of cells that constitute an intrinsic void as the centre of that void.
This is called the \enquote*{barycentre} in the {\revolvername} code, but is mathematically the barycentre only if the void is assumed to be filled with a uniform density fluid \citep[][\SSS{}1]{PeperRoukema2020}.

\subsubsection{Photometric void detection} \label{s-method-photo-voids}

\begin{table}
  \centering
  \caption{Parameters used in our {\twD} void detection algorithm in Eqs~\eqref{e-crit-supernoise-increase} and \eqref{e-crit-mean}.
    \label{Void-crit-table}}
  $\begin{array}{l c c}
    \hline
    & f_{\mathrm{std}} & f_{\mathrm{mean}} \\
    \hline
    \Sigma
    & {\VoidStddevFracCritvalue} & {\VoidDiffFracCritvalue} \rule{0ex}{3ex} \\
    \theta
    & {\VoidStddevFracCritThetavalue} & {\VoidDiffFracCritThetavalue} \\
    \sigma
    & {\VoidStddevFracCritSigmavalue} & {\VoidDiffFracCritSigmavalue} \\
    \hline
  \end{array}$
\end{table}

We detect voids from variables in the sky plane that are, in principle, observationally measurable in photometric surveys: the surface overdensity $\Sigma$, and three geometric-optics related parameters.
We propose the following heuristically derived algorithm for detecting a projected void in a map of the surface overdensity $\Sigma$ (defined below in \SSS\ref{s-meth-mass-def}) or the \citet{Sachs1961} expansion $\theta$ or shear $\sigma$ (optical scalars, defined below in \SSS\ref{s-meth-opt-scalars}).
Our algorithm for detecting voids from maps of the weak-lensing shear $\gamma$ is similar, but differs in the ways that are described below in \SSS\ref{s-meth-lensing-gamma}.

In contrast to the case for overdense extragalactic objects, we expect the (azimuthally averaged) radial density profile of a void in the sky plane, where the radius is $r := \sqrt{(x-x_0)^2 + (z-z_0)^2}$ in our $(x,z)$ grid for an object centre $(x_0,z_0)$, to have its lowest values in the centre of the void and a sharp maximum at the void's edge.
The Sachs expansion $\theta$ and shear $\sigma$ should also have a minimum at the centre of a void and a maximum at the edge.
While projection effects for overdense structures are a perennial problem in astronomy (e.g. for determining whether a galaxy group is dynamically real or a chance projection), the projection effects can be expected to be much worse for voids, since voids dominate the volume of the Universe, implying stronger overlaps.
In contrast to overdense objects, spectroscopic redshift determination for the rare galaxies in voids is unlikely to be effective in separating a chance projection of voids from a void that is real in three spatial dimensions, and is likely to be a statistically unstable way of dynamically characterising a void.
Moreover, voids, in general, are not perfectly spherical objects, making detection via templates unlikely to be easy.
Nevertheless, projection along the line of sight should provide a modest effect of symmetrisation, and by appropriately averaging, we hypothesize that detection is feasible.

We define an isotropised (azimuthally averaged) variable $\overline{X}$, i.e. the average on a circle in the $(x,y \equiv \chi_{\mathrm{O}},z)$ grid plane, at radius $r$ and centred on a pixel $j$, i.e.
\begin{equation} \label{eq-circ-avg}
\overline{X}_j(r) = (2\rmpi)^{-1} {\int_0^{2\varphi} X(r, \varphi) \,\diffd\varphi} \,,
\end{equation}
where $\varphi$ is the angle around the circle centred on position $j$ in the grid and $X$ is either the surface overdensity $\Sigma$, or one of the optical scalars $\theta$ or $\sigma$ (for the weak lensing shear $\gamma$, see Eq.~\eqref{eq-weak-lens-gamma} below).
To estimate $\overline{X}_j(r)$, we sample the grid values at even intervals around a circle of radius $r$, with intervals that give at least one value per Mpc/$h$, we smooth the values with a third-order Savitzky--Golay filter \citep{SavGol1964}, and integrate.

We also define a disc-averaged profile $\doublemeanX$ on the disc internal to a given radius (not weighted by the radius) by integrating Eq.~\eqref{eq-circ-avg} and appropriately normalising, i.e.
\begin{equation}
  \doublemeanX(r) = \frac{\int_0^r \int_0^{2\rmpi} X(r^\prime, \varphi)\, \mathrm{d}\,\varphi \mathrm{d}r^\prime}{(2\rmpi)^{-1}\int_0^r \int_0^{2\rmpi} \mathrm{d}\varphi\,\mathrm{d}r^\prime} = \frac{ \int_0^r \barX(r^\prime)\,\mathrm{d}r^\prime}{r} \,,
  \label{eq-disc-doublemean}
\end{equation}
and a disc-averaged absolute slope,
\begin{align}
  \meanabsslopeX(r_i)
  &= \frac{\sum_{i'<i} \left\lvert\diffd \barX/\diffd r\right\rvert(r_{i'})}{i-1}
\,,
  \label{eq-mean-slope}
\end{align}
where the index $i$ indicates radial discretisation in estimating $\barX$ and $\barX' := \diffd \barX/\diffd r$.

Apart from the case of the weak-lensing shear $\gamma$ (\SSS\ref{s-meth-lensing-gamma}), we expect $\overline{X}$ (and thus $\doublemeanX$) to increase monotonically from the centre at $r=0$ outwards as $r$ increases, though the projection against other voids, voids' asphericity, and noise will make this monotonicity difficult to detect.
For each pixel $j$ in our sky plane, we define a heuristic selection criterion $\eta_X$ motivated by the expected monotonicity as follows.
\begin{list}{(\roman{enumi})}{\usecounter{enumi}}
\item Ignore each pixel $j$ with $ X(r=0) > X_{\mathrm{median}}$, where $X_{\mathrm{median}}$ is the median over all pixels in the sky plane.
  The motivation is that for any $X$, pixels with $ X(r=0) > X_{\mathrm{median}}$ are unlikely to correspond to the centre of a void.
  The projected or ray-traced variables $X$ should have their minima at voids' centres.
  This step should remove many pixels unlikely to be void centres.
  \label{algo-item-rm-high-pixels-j}
\item For a given pixel $j$, for each radial distance $r_i$ from the pixel, calculate the circular average $\barX$ as in Eq.~\eqref{eq-circ-avg}, where for simplicity we omit the subscript $j$.
  We set the interval in $r_i$ to be smaller than 1~Mpc/$h$ in order to be sensitive to small-scale structure.
  \label{algo-item-radial-Xbar}
\item A persistently positive strong positive slope in $\barX$ is detected as follows.
  Find $i_1$, the first radial position $i$, with respect to pixel $j$,
    where all three of the conditions
    \begin{align}
\barX^\prime(r_i) & > \meanabsslopeX(r_i)
      \label{e-crit-increase} \\
\barX(r_i) - \barX(r_{i-1}) & >
      f_{\mathrm{std}} \left\langle \left(\barX(r_{i'}) - \doublemeanX(r_{i'})\right)^2 \right\rangle_{i'<i}^{1/2}
      \label{e-crit-supernoise-increase} \\
\barX(r_i) &> f_{\mathrm{mean}} \,\doublemeanX(r_i)
      \label{e-crit-mean}
    \end{align}
    are satisfied over four successive steps $i-3, i-2, i-1, i$,
    where $\langle \cdot \rangle$ is the mean and $f_{\mathrm{std}}$ and $f_{\mathrm{mean}}$ are heuristically chosen fractions.
The aim of criteria \eqref{e-crit-increase} and \eqref{e-crit-supernoise-increase} is to find a range of radii where the slope $\barX'$ has a stable and significant increase, i.e. where positive second derivatives $\barX'' > 0$ are numerically persistent.
    Criterion \eqref{e-crit-mean} aims to also require the slope $\barX'$ to be sufficiently positive.
    The values adopted for $f_{\mathrm{std}}$ and $f_{\mathrm{mean}}$ are given in Table~\ref{Void-crit-table}.
    \label{algo-item-strong-pos-slope}
  \item Find $i_2$, the first local maximum in $\barX$ for $i > i_1$, i.e., the first local maximum after the persistently strong positive slope condition that determines $i_1$.
    \label{algo-item-first-local-max}
  \item Define an initial void selection criterion $\eta^0_X(j) := 1/r_{i_2}$.
    The radius $r_{i_2}$ is the estimated radius of the candidate void.
    \label{algo-item-define-eta-selection}
  \item Steps \ref{algo-item-radial-Xbar}--\ref{algo-item-define-eta-selection} are carried out for all pixels $j$ accepted in step \ref{algo-item-rm-high-pixels-j} (with \ref{algo-item-strong-pos-slope}, \ref{algo-item-first-local-max} modified in the case of $\gamma$; see \SSS\ref{s-meth-lensing-gamma}).
    In order to cope with the very noisy data, we define a smoothed selection criterion for further use, $\eta_X(j)$, as a low-pass triangular filter (weighted mean) of the ${\MassDefNAveragePixelvalue}$ $\eta^0_X(j)$ values in a ${\NAveragevalue}\times{\NAveragevalue}$ grid of pixels centred on pixel $j$\footnote{This step interprets the box as an isolated box, not $T^2$, and sets $\eta_X(j)$ near the borders of the box to a high value to prevent finding minima there.}.
\end{list}
We then find all local minima of the selection criterion $\eta_X(j)$ over the pixels $j$ as follows.
\begin{list}{(\roman{enumi})}{\usecounter{enumi}\addtocounter{enumi}{6}}
\item We select a void centred at pixel $j$ if it dominates its local region in the sense that $\eta_X(j) < \eta_X(k)$ where $k$ indexes pixels in a square grid centred on pixel $j$ and extending $\NNeighboursvalue$ pixels in each of the $\pm x$ and $\pm z$ directions.
  As an extra credibility criterion, selection of a void is only accepted if $\eta_X(j) < \ToleranceFracMeanvalue \sum_k \eta_X(k)/\sum_k 1$, where $k$ indexes all pixels in the map.
  \label{algo-item-min-local-eta}
\end{list}
This algorithm results in a list of projected voids with centres $j$ and radii $r_{i_2}(j)$ that represent the largest locally credible voids.

\begin{list}{(\roman{enumi})}{\usecounter{enumi}\addtocounter{enumi}{7}}
\item  To avoid cases where a single genuine void is misidentified as two slightly offset voids, we check if two or more centres are closer to one another than
  \begin{align}
    \min\left\{ R(j_1)/4, R(j_2)/4, \MergeDistancevalue\,\mathrm{Mpc}/h \right\}\,.
  \end{align}
  In these cases, we merge these voids into a single void.
  The new centre and radius of the merged void are defined as the mean of the centres and radii of the unmerged voids.
\end{list}

\subsection{Matches to intrinsic voids} \label{s-method-matches}
To quantify whether the photometric detection of voids -- {\twD} voids -- successfully finds the intrinsic {\thrD} voids, we first define a heuristically motivated matching criterion to find the best matches, and then compare the set of best matches to an equivalent set of best matches when the list of {\twD} void parameters is generated randomly (positions) or randomly shuffled (radii).
This aims at answering two different questions: (i) given a detected set of {\twD} voids, are these better than a random set of {\twD} voids at revealing true {\thrD} voids? (ii) given a set of intrinsic {\thrD} voids, do the detected {\twD} voids better match these (numerically) real voids better than a random set of {\twD} voids would?
The former question is interesting for observational detection of {\thrD} voids from photometric or other geometric-optics data; the latter is interesting for using spectroscopically defined {\thrD} voids to motivate searches for gravitational lensing by voids.

\subsubsection{Best matches and median sky-plane separaration $\mu_{x,z}$}

We define the probability of the $i$th {\twD} void being a match to the $j$th {\thrD} (watershed) void by first defining the probabilities that the $x$ and $z$ positions are close in the $T^2$ sense and that the radii are logarithmically close.
We set a cumulative Gaussian probability that the $x$ or $z$ positions for variable $X \in \{\Sigma, \gamma, \theta, \sigma\}$ are closer to each other than the estimated values,
\begin{align}
  P^X_{x_{i,j}} &= 1 - \erf \frac{d\left(x^X_i, x^X_j\right)}{\sqrt{2}\,\sigma_x} \nonumber \\
  P^X_{z_{i,j}} &= 1 - \erf \frac{d\left(z^X_i, z^X_j\right)}{\sqrt{2}\,\sigma_z} \,,
  \label{e-xz-search-prob}
\end{align}
where $\erf$ is the error function, $d(.,.)$ is the $T^2$ minimum $x$ or $z$ distance,
and $\sigma_x = \sigma_z = {\StdXzVoidProbvalue}\,\mathrm{Mpc}/h$.
Similarly, we set
\begin{align}
  P^X_{R_{i,j}} &= 1 - \erf \frac{\left\lvert\log_{10}\left(R^X_i/R^X_j\right)\right\rvert}{\sqrt{2}\,\sigma_{\log_{10}R}} \label{e-R-search-prob}
  \,,
\end{align}
where $\sigma_{\log_{10}R} = {\StdRVoidProbvalue}$~dex \citep{Allen51dex}.
These are assumed, for simplicity, to be independent probabilities, giving a heuristic overall probability that the $i$th {\twD} void matches the $j$th {\thrD} void
\begin{align}
  P^X_{i,j} &= P^X_{x_{i,j}} \, P^X_{z_{i,j}} \, P^X_{R_{i,j}} \,.
  \label{e-xzR-search-prob}
\end{align}

For question (i) (\SSS\ref{s-method-matches}), given a {\twD} void $i$, we find the {\thrD} void $j$ with the highest matching probability $P^X_{i,j}$, for detector variable $X$.
This does not exclude the possibility that two different {\twD} voids best identify with the same {\thrD} void.
For a set of $N_{2D}$ detected {\twD} voids, this gives us a matched set of $N_{2D}$ objects, which have both {\twD} and {\thrD} sky position and radius information, presumed to match.

For each object in this set, we calculate the $T^2$ distance between the {\twD} and {\thrD} $(x,z)$ positions and from the distribution of these values, calculate $\mu_{x,z}(3D|2D)$, the median distance for a {\thrD} match given a {\twD} match.
In calculating this median, in cases where a single {\thrD} void is the best match for two or more {\twD} voids, we only consider the match in which $P^X_{i,j}$ is maximised.

For question (ii), given a {\thrD} void $i$, we find the {\twD} void $j$ with the highest matching probability $P_{i,j}$.
Again, this does not exclude the possibility that two different {\thrD} voids best identify with the same {\twD} void.
In practice, since we find fewer {\twD} voids to {\thrD} voids, there are necessarily cases where multiple {\thrD} voids identify with a single {\twD} void.
For a set of $N_{3D}$ detected {\thrD} voids, this gives us a matched set of $N_{3D}$ objects, which have both {\twD} and {\thrD} sky position and radius information, presumed to match.

For each object in this set, we calculate the $T^2$ distance between the {\twD} and {\thrD} $(x,z)$ positions and infer $\mu_{x,z}(2D|3D)$, the median distance for a {\twD} match given a {\thrD} match, again using the highest $P^X_{i,j}$ to reduce non--one-to-one matches.

\subsubsection{Comparison to matches for random {\twD} voids} \label{s-prob-xz-prob-R}

We generate a Monte Carlo simulation of {\twD} voids by choosing $N_{2D}$ pairs ($x,z$) from a uniform random distribution within the $x$ and $z$ ranges of the {\twD} grid.
For both questions (i) and (ii) (separately), for each Monte Carlo simulation, we find matched sets using the same algorithm as above.

To answer question (i) we estimate $\mu_{x,z}(3D|2D)$ for a given detector $X$, for both the original matched set and for each of the simulated matched sets.
This yields $P^X_{x,z}(3D|2D)$, the frequentist probability that the original $\mu_{x,z}(3D|2D)$ is less than $\mu_{x,z}(3D|2D)$ from the Monte Carlo simulations.
In other words, $P^X_{x,z}(3D|2D)$ is the probability that, given the {\twD} voids, the matches with {\thrD} voids in the sky plane are no better than those drawn from a Monte Carlo simulation.

Similarly, for question (ii), the frequency with which the original $\mu_{x,z}(2D|3D)$ is less than the values of $\mu_{x,z}(2D|3D)$ from the simulations yields $P^X_{x,z}(3D|2D)$, the probability that, given the {\thrD} voids, the matches with {\twD} voids in the sky plane are no better than those drawn from a Monte Carlo simulation of sky positions.

Cases where {\thrD} voids are concentric or approximately overlap in projection will yield only a single {\twD} void using our algorithm, and are likely to make estimation of the radii difficult.
The range of values of the radii are not as conveniently constrained as the $(x,z)$ centres of the voids.
Rather than choosing an arbitrary range for a Monte Carlo simulation, we use a non-parametric method.
We define $P^X_R(3D|2D)$ as the two-sided probability that the Spearman $\rho$ rank correlation coefficient \citep{Spearman1904} of the matched set of $N_{2D}$ values $R^X_i$ and $R^X_{j(i)}$ is stronger (positive or negative) than what it would be for a set of paired values where one set is randomly permuted.

Similarly, we define $P^X_R(2D|3D)$ as the two-sided probability that the Spearman $\rho$ ranking coefficient of the matched set of $N_{3D}$ values $R^X_i$ and $R^X_{j(i)}$ is stronger (positive or negative) than it would be under random permutations.

\subsection{Detector variables $\Sigma, \gamma, \theta, \sigma$} \label{s-model}

Here we define and describe our detector variables $X \in \{\Sigma, \gamma, \theta, \sigma\}$, where here we write the generic forms of these variables for simplicity; the more specific forms are given below.
These detector variables can, in principle, be derived from a photometric map, given some minimal assumptions, such as a mass-to-light ratio in the case of the surface overdensity $\Sigma$, or statistically isotropic distributions of galaxy shape parameters in the case of the other three parameters.
We derive each of these from the particle distribution, not from the galaxy distribution.
We include $\Sigma$ since apart from requiring a mass-to-light ratio assumption, it is the simplest to derive from a photometric map.

\subsubsection{Surface overdensity $\Sigma$} \label{s-meth-mass-def}
We calculate the surface overdensity by integrating the overdensity $\rho - \bar{\rho}$ along the line of sight, neglecting temporal evolution.
(Temporal evolution is taken into account with the optical scalar modelling; see \SSS\ref{s-meth-opt-scalars} below.)
Densities are constructed for each particle using a Voronoi tessellation followed by linear interpolation.
For a flat model, the surface overdensity in direction $\hat{n}$ is
\begin{equation}
  \Sigma(\hat{n}) = \int_{\chi_\mathrm{min}}^{\chi_\mathrm{max}} (\rho(\hat{\chi},\Omega) - \bar{\rho} ) \mathrm{d}\chi ,
\end{equation}
where $\chi_\mathrm{min} = \chi(z_\mathrm{O}) - 0.95 L_\mathrm{box}/2$ and $\chi_\mathrm{max} = \chi(z_\mathrm{O}) + 0.95 L_\mathrm{box}/2$.
The 0.95 factor neglects the 5\% front and back parts of the box to minimise edge effects (\SSS\ref{s-method-geometry}).
Since we expect the surface overdensity $\Sigma$ to be negative in a void, we aim to detect it in places where it is physically a surface underdensity, i.e., a projected mass deficit.

\paragraph{Detection strategy with $\Sigma$}
Voids typically have strong underdensities in their interior, so the {\twD} projection of a void should still show a strong underdensity in the interior after projection.
Thus, we search for local minima in $\Sigma$.

The projection of foreground and background voids and their walls (in reality, clusters, filaments and walls) will, to some degree, obscure this search.
To the extent that the obscuration can be statistically neglected or removed, the azimuthally averaged radial profile $\barSigma$ (Eq.~\eqref{eq-circ-avg}) should show a slow increase in $\barSigma$ from the centre of a projected void outwards until it nears the (projected) wall, when a rapid increase should occur, followed by a drop as $\barSigma$ asymptotes to the mean of the environment surrounding the void.
Thus, a local maximum in $\barSigma$ at the wall that surrounds the void should be sought.
We expect that larger voids should yield clearer signals.

Some pixels, likely containing galaxy clusters or projections of galaxy filaments, were found to be highly overdense, misleading our algorithm's search for void walls because of these overdense pixels' strong influence on $\barSigma$.
To reduce the influence of these extreme pixels, prior to step \ref{algo-item-rm-high-pixels-j}, we truncate $\Sigma$ values at the $\CosmicRayPercentilevalue$th percentile of their distribution.

Substructures of overdensities exist inside of voids, similar to the larger scale overdensities of the cosmic web, but traced by dark matter haloes of much lower mass \citep{Gottloeber2003}.
These substructures contribute another obscuring factor that should weaken our proposed detection algorithm (\SSS ~\ref{s-method-voids}) using $\Sigma$.

\subsubsection{Weak-lensing tangential shear $\bargamma$} \label{s-meth-weak-lensing}

Weak gravitational lensing information is typically extracted from observations by using the distortion of observed images that is induced by cosmological inhomogeneity, with the aim of tracing the spatial distribution of dark matter.
We follow the mathematical descriptions of \citet{BartelmannSchneider01,Krause2013} and \citet{Kilbinger2015WeakLensing}.
We derive the parameters of this idealised model from the surface overdensity calculated in our cosmological simulation.

We represent the lens plane \citep[][fig.~11]{BartelmannSchneider01} with two orthogonal spatial directions with indices $a$ and $b$; the direction of propagation of the light bundle as it would arrive at the observer from the source if unlensed, represented as a vector in the lens plane, $\Theta_S$; and the direction at which the light bundle reaches the observer after lensing, again a vector in the lens plane, $\Theta_O$.
A matrix to convert from the observed directions to the original source directions, the \enquote*{deformation matrix} \citep{Hossen2022ringing}, $A$ is defined as the Jacobian
\begin{equation}
  \Theta^a_S = A^a_b \Theta^b_O \,.
\end{equation}
Assuming that the rotation of the image vanishes, the deformation matrix can be decomposed into the shear $\gamma$ and the convergence $\kappa$ \citep[][eq.~(3.11)]{BartelmannSchneider01}:
\begin{equation}
  A= \left( \begin{matrix}
    1 - \kappa - \gamma_1 & \gamma_2 \\
    \gamma_2 & 1 - \kappa + \gamma_1
    \end{matrix}
    \right)\,.
  \end{equation}
The convergence $\kappa$ at a generic position in the sky plane can be evaluated as
$\kappa(\hat{n}) = \tfrac{\Sigma(\hat{n})}{\Sigma_{\mathrm{crit}}}$,
where $\Sigma_{\mathrm{crit}} = \frac{c^2}{4 \rmpi G} \frac{D_{\mathrm{OS}}}{D_{\mathrm{OL}} D_{\mathrm{LS}}}$
\citep[][eq.~(3.7)]{BartelmannSchneider01}.
The values $D_{\mathrm{XX}}$ are the angular diameter distances between the observer (O), the lens (L) and the source (S).
We do not attempt void detection with $\kappa$, as the result would be equivalent to using $\Sigma$.

For our detection strategy, we use $\barSigma(r,\hat{n})$, the isotropised (ring averaged) form of $\Sigma$ (Eq.~\eqref{eq-circ-avg}) with respect to a given centre $\hat{n}$ of a possible void, and subtract it from the surface overdensity averaged within a disc centred on $\hat{n}$, yielding
\begin{align}
  \Delta \Sigma(r,\hat{n}) := \tildeSigma(r,\hat{n}) - \barSigma(r,\hat{n})
  \,,
  \label{e-defn-DeltaSigma}
\end{align}
where $\tildeSigma(r,\hat{n})$ is defined in Eq.~\eqref{eq-disc-avg-sigma}.
As in \citet[][eqs~(4), (5)]{Krause2013} and also derived in \citet[][eqs~(40)--(47)]{Kilbinger2015WeakLensing}, the mean of the tangential component of the shear internal to a ring at $r$ can then be evaluated as
\begin{align}
  \bargamma(r,\hat{n}) = \frac{\Delta \Sigma(r,\hat{n})}{\Sigma_{\mathrm{crit}}} \,,
  \label{eq-weak-lens-gamma}
\end{align}
where we leave the weak dependence of $\Sigma_{\mathrm{crit}}$ on $r$ and $\hat{n}$ implicit.
The disc average $\tildeSigma$ calculated by integrating \eqref{eq-circ-avg} over the radius, using the usual weighting and now leaving the centre $\hat{n}$ implicit, is
\begin{equation}
  \tildeSigma(r)=\frac{\int_0^r \int_0^{2\rmpi} \Sigma(r^\prime, \varphi) r^\prime \mathrm{d}\varphi \mathrm{d}r^\prime}{\int_0^r \int_0^{2\rmpi} r^\prime \mathrm{d}\varphi\mathrm{d}r^\prime} = \frac{ 2\rmpi \int_0^r \barSigma(r^\prime) r^\prime\,\mathrm{d}r^\prime}{\rmpi r^2}
  \,.
  \label{eq-disc-avg-sigma}
\end{equation}

\paragraph{Detection strategy with $\bargamma$} \label{s-meth-lensing-gamma}
By definition, $\bargamma(r)$ should be close to zero at $r=0$, the centre of a void, and should decrease to a sharp minimum where $r$ is the radius of the void's (statistical) wall.
At greater radii, both the azimuthally averaged surface overdensity $\barSigma$ and the disc-averaged $\tildeSigma$ should approach zero, so $\bargamma(r)$ should also increase up to zero.
The minimum in $\bargamma(r)$ should reveal the edge of the void.

Since this qualitative behaviour of $\bargamma(r)$ differs from the other detector variables considered, we modify steps \ref{algo-item-strong-pos-slope} and \ref{algo-item-first-local-max} of the algorithm of \SSS\ref{s-method-photo-voids} as follows.

Since $\bargamma$ and $\bargamma'$ calculated according to \eqref{eq-circ-avg} are noisy, we apply extra smoothing, replacing $\bargamma(r_i)$ and $\bargamma'(r_i)$ by $\langle \bargamma(r_i) \rangle_{\{\max(0,{i-3}), \ldots, {i+3}\}}$ and $\langle \bargamma'(r_i) \rangle_{\{\max(0,{i-3}), \ldots, {i+3}\}}$, respectively.
This smoothing reduces the role of local fluctuations in the dark matter distribution.

\begin{list}{(\roman{enumi}')}{\usecounter{enumi}\addtocounter{enumi}{2}}
\item After this smoothing, we search for the radial distance where $\bargamma$ starts dropping sharply, i.e. the index $i_1$ is the first value $i$ where $\bargamma(r_i) < \doublemeangamma(r_i)$.
\item The radial distance just past the wall is sought as the radial distance where $\bargamma$ increases sharply, i.e. the index $i_2$ is the first value $i > i_1$ where $\bargamma(r_i) > \doublemeangamma(r_i)$.
\end{list}
\begin{list}{hardwired}{}
  \item[(iv'.1)]
    In addition, to remove choices of a void centre where the best \enquote*{wall} found this way has a weak density contrast, we dismiss the candidate detection if $ \left\lvert\doublemeangamma\right\rvert / \max\left(\left\lvert\bargamma\right\rvert\right)  < 0.1 $.
    For patterns in $\bargamma$ that have almost no significant features, this criterion avoids interpreting a nearly flat curve $\bargamma(r_i)$ as a candidate void.
\end{list}

If both $r_{i_1}$ and $r_{i_2}$ are detected, then we continue to step \ref{algo-item-define-eta-selection} as above (\SSS\ref{s-method-photo-voids}).
Even if pixel $j$ is correctly centred on a void's centre, this algorithm for $\bargamma$ can fail to detect $r_{i_2}$ if the (projected) environment just outside the void's wall includes strong fluctuations.
In the case of failure to detect $r_{i_2}$, the pixel is considered invalid at step \ref{algo-item-define-eta-selection} and dropped from further consideration.

\subsubsection{Optical scalars $\theta$ and $\lvert\sigma\rvert$} \label{s-meth-opt-scalars}
We calculate optical scalars following \citet{Sasaki1993}.
In principle, these should model the real Universe more accurately than the weak-lensing approach described above, since fewer assumptions are required.
In the Newtonian approximation, the Ricci tensor can be written as
\begin{equation}
  R_{00} \approx 8 \rmpi G \rho \omega^2
  \label{e-Ricci00}
\end{equation}
and the Weyl tensor components of interest are
\begin{align}
  C_{A0B0} &\approx (2\Phi_{;AB} - \delta_{AB} \Phi^{;C}_{;C})\omega^2 \\
  &= (2\Phi_{;\mu \nu} e_A^\mu e_B^\nu - \delta_{AB} \delta^{CD} \Phi_{;\mu \nu} e_C^\mu e_D^\nu) \omega^2 \,,
    \label{e-Weyl_A0B0}
\end{align}
\citep[][eqs~(3.22), (3.21)]{Sasaki1993}
where $G$ is the gravitational constant; space and time units are related by $c=1$; $\rho$ is density; $\Phi$ is the gravitational potential; $\omega = -k_\mu u^\mu = 1+z_\mathrm{redsh}$, for an observer four-velocity $u^\mu$, light propagation one-form $k_\mu$, and redshift $z_\mathrm{redsh}$; $\{e_A, e_B\}$ or $\{e_C, e_D\}$ are a pair of dyad basis vectors; and $\delta_{AB}$ is the Kronecker delta ($\delta_{AB} = 1$ if $A=B$, $\delta_{AB} = 0$ if $A\ne B$).
The dyad basis vectors $e_A, e_B$ span the spacelike 2-plane that is orthogonal to the spatial path of the light ray that points from the observer towards the direction of a cell of the grid.
We use the Gram--Schmidt method to construct $e_A$ and $e_B$.

The optical scalars -- the expansion $\theta$ (real) and the shear $\sigma$ (complex) -- are related to each other and the Weyl tensor (Eq.~\eqref{e-Weyl_A0B0}) via the coupled pair of differential equations
\begin{equation}
  \frac{\mathrm{d}}{\mathrm{d}\varv} \theta = -R_{00} -2 |\sigma|^2 - \frac{1}{2} \theta^2
  \label{eq-optical-scalars-theta}
\end{equation}
and
\begin{equation}
  \frac{\mathrm{d}}{\mathrm{d}\varv} \sigma = -(C_{1010} + \mathrm{i}\, C_{1020}) - \sigma \theta \,.
  \label{eq-optical-scalars-sigma}
\end{equation}
For a visualisation of the effect of $\theta$ and $\sigma$ on a light bundle we refer to \citet[][Fig.~4]{Sasaki1993}, where $\sigma = \sigma_{+} + \mathrm{i}\, \sigma_{\times}$.
The dependence of the optical scalars $\theta$ and $\sigma$ to the usual weak-lensing parameters is given in eqs~(41)--(43) of \citet{Clarkson12lensingbias}.

\paragraph{Detection strategy with $\theta$ and $\lvert\sigma\rvert$} \label{s-meth-lensing-opt-scalars}

Both the expansion $\theta$ and the modulus of the shear,
\begin{align}
  \lvert\sigma\rvert &= \sqrt{\mathrm{Re}(\sigma)^2 + \mathrm{Im}(\sigma)^2}
  \label{e-sigma-modulus}
\end{align}
should be closely related to the surface overdensity, since integrations along paths approximately (spatially) orthogonal to the lens plane are performed in all three cases.
However, these are not exactly analogous.
Not only are these distinct physical quantities, but the overdensity integral is performed parallel to the $y$ axis, while for each pixel in our two-dimensional grid plane, we estimate $\theta$ and $\lvert\sigma\rvert$ along a spatially straight path from the observer through the pixel, i.e. only approximately parallel to the $y$ axis.

\begin{table}
  \centering
  \caption{Numbers of intrinsic {\thrD} voids detected with {\revolvername}, $N_{3D}$, and in the {\twD} grid,
    $N_{2D}^\Sigma$,
    $N_{2D}^\gamma$,
    $N_{2D}^\theta$, and
    $N_{2D}^\sigma$,
    using the surface overdensity $\Sigma$,
    the weak-lensing shear $\bargamma$,
    the Sachs expansion $\theta$, and
    the modulus of the Sachs shear $|\sigma|$, respectively.
    \label{t-N-detections}}
  $\begin{array}{l c c c c}
    \hline
    \rule{0ex}{2.7ex} N_{3D} &
    N_{2D}^\Sigma &
    N_{2D}^\gamma &
    N_{2D}^\theta &
    N_{2D}^\sigma
    \rule[-1.2ex]{0ex}{0ex} \\
    \hline
    \MassDefNVoidRevolvervalue &
    \MassDefNVoidvalue &
    \MassDefNVoidGammavalue &
    \OSNVoidExpvalue &
    \OSNVoidSigmavalue \\
    \hline
  \end{array}$
\end{table}

\begin{table}
  \centering
  \caption{Probability that the matches between {\thrD} and {\twD} voids for detector variable $X$ are no better than those of randomly generated {\twD} voids, $P^X_{xz}(3D|2D)$ when given {\twD} voids; or $P^X_{xz}(2D|3D)$ when given {\thrD} voids; and probability that the Spearman rank correlation coefficient for the radii of matched {\thrD} and {\twD} voids for detector variable $X$ could be that of a set of randomly paired values, $P^X_R(3D|2D)$ when given a {\twD} void; and $P^X_R(2D|3D)$ when given a {\thrD} void. See \SSS\protect\ref{s-prob-xz-prob-R}.
    Plain-text version available at \href{\projectzenodofilesbase/void_match_analysis.dat}{\projectzenodoid/void\_match\_analysis.dat}.
    \label{t-match-significance}}
  $\begin{array}{l c c c c}
    \hline
    \rule{0ex}{2.7ex} X & P_{x,z}(3D|2D) & P_{x,z}(2D|3D) & P_R(3D|2D) & P_R(2D|3D) \\
    \hline
    \rule{0ex}{3ex} \VoidMatchSignificanceTable
    \hline
  \end{array}$
\end{table}

In practice, initial numerical exploration shows that $\theta$ and $\lvert\sigma\rvert$ behave qualitatively like $\Sigma$, in that they start from a low value at the centre of a void and increase to a sharp maximum at a void boundary.
Thus, we use the same search algorithm for finding voids in maps of $\theta$ and $\lvert\sigma\rvert$ as indicated above in \SSS\ref{s-method-photo-voids}, with slightly adjusted parameters (Table~\ref{Void-crit-table}).
While qualitatively similar in numerical terms, the physical meanings of these parameters differ.
The optical scalars $\theta$ and $\lvert\sigma\rvert$, if derived from observations, represent the underlying matter distribution with no dependence on observed luminosity and without the simplifying assumptions of the weak-lensing approximation.
In particular, the weak-lensing shear $\bargamma$ (Eq.~\eqref{eq-weak-lens-gamma}) is an average defined with respect to a hypothesized void centre, while $\theta$ and $\lvert\sigma\rvert$ provide maps prior to assumptions about void centres.

\section{Results} \label{s-results}

\subsection{Simulation} \label{s-results-simulation}

\newcommand\TemporaryForDraftOnly{
\begin{figure}
  \includegraphics[width=0.9\columnwidth]{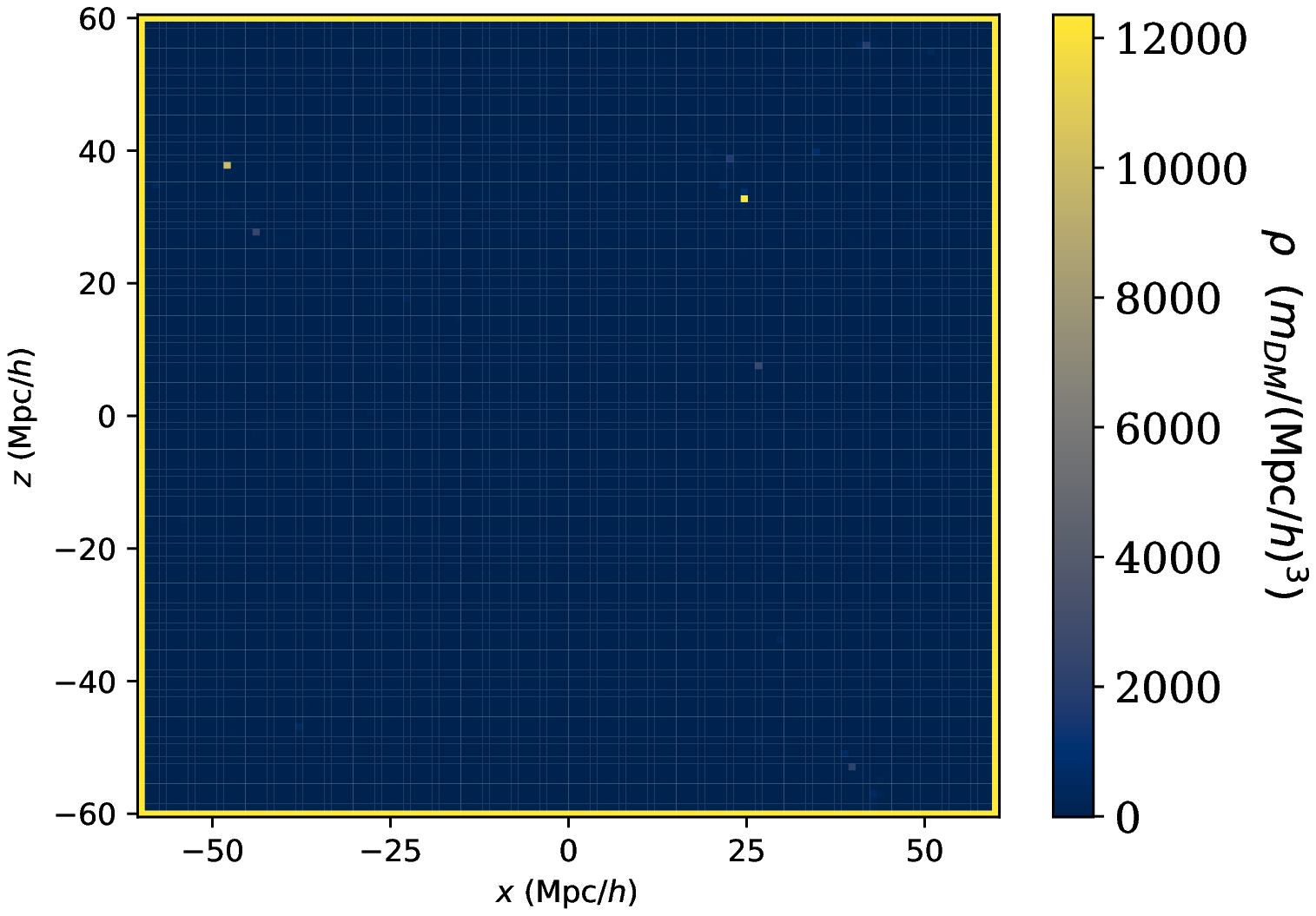}
  \caption{
    The densities of the slice with $y_{\mathrm{S}}=0$.
    The void is surrounded by an overdense region.
    Non-void galaxies, classified by Revolver, are shown in red; void galaxies are shown in blue.
\label{fig-dens}}
\end{figure}
\begin{figure}
  \includegraphics[width=0.9\columnwidth]{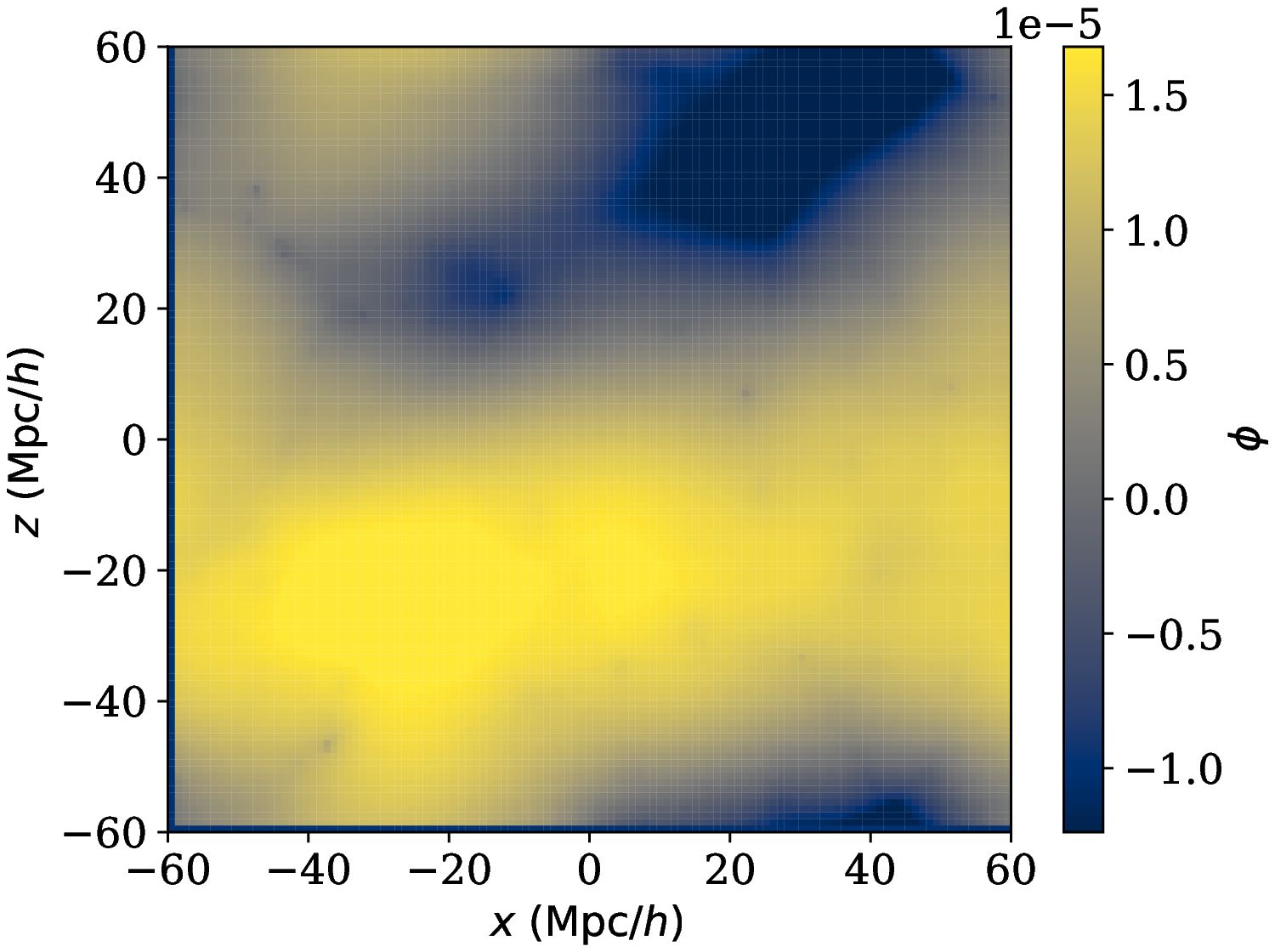}
  \caption{
    The potential of the slice with $y_{\mathrm{S}}=0$.
    As expected we see a clear maximum of the potential inside the void.
\label{fig-pot}}
\end{figure}

Figures~\ref{fig-dens} and \ref{fig-pot} show the densities and the gravitational potential for a slice with $y_{\mathrm{S}}=0$ for our realisation.
The gravitational potential of the larger voids dominates the potential map of the simulation.
\textbf{\textcolor{red}{These figures are not really interesting as they neither show anything new nor any result. The figures are only there t visualise these two key properties while we write the paper.}}
}

We performed an {\it ab initio} simulation and detected voids as described above.
As indicated in Table~\ref{t-N-detections}, we detected $N_{3D} = {\MassDefNVoidRevolvervalue}$ voids in the galaxy population with the watershed mechanism, and smaller numbers of {\twD} voids using $\Sigma$, $\bargamma$, $\theta$, and $\lvert\sigma\rvert$ from the projected density distribution and by ray-tracing through the evolving gravitational potential $\Phi$.

Table~\ref{t-match-significance} shows the probabilities, defined in \SSS\ref{s-prob-xz-prob-R}, that quantify the significance of:
(i) a detected {\twD} void revealing the existence of an intrinsic {\thrD} void via its sky plane position or radius, $P^X_{x,z}(3D|2D)$ or $P^X_{R}(3D|2D)$, respectively,
and (ii) an intrinsic {\thrD} void implying that its {\twD} projection is detectable, $P^X_{x,z}(2D|3D)$ or $P^X_{R}(2D|3D)$.
In each case, these represent the probability that the estimated correspondence between the populations could occur by chance, given prior information on the number of {\twD} voids (for positions) or non-parametrically (for radii).

\begin{figure}
  \centering
  \includegraphics[width=0.95\columnwidth]{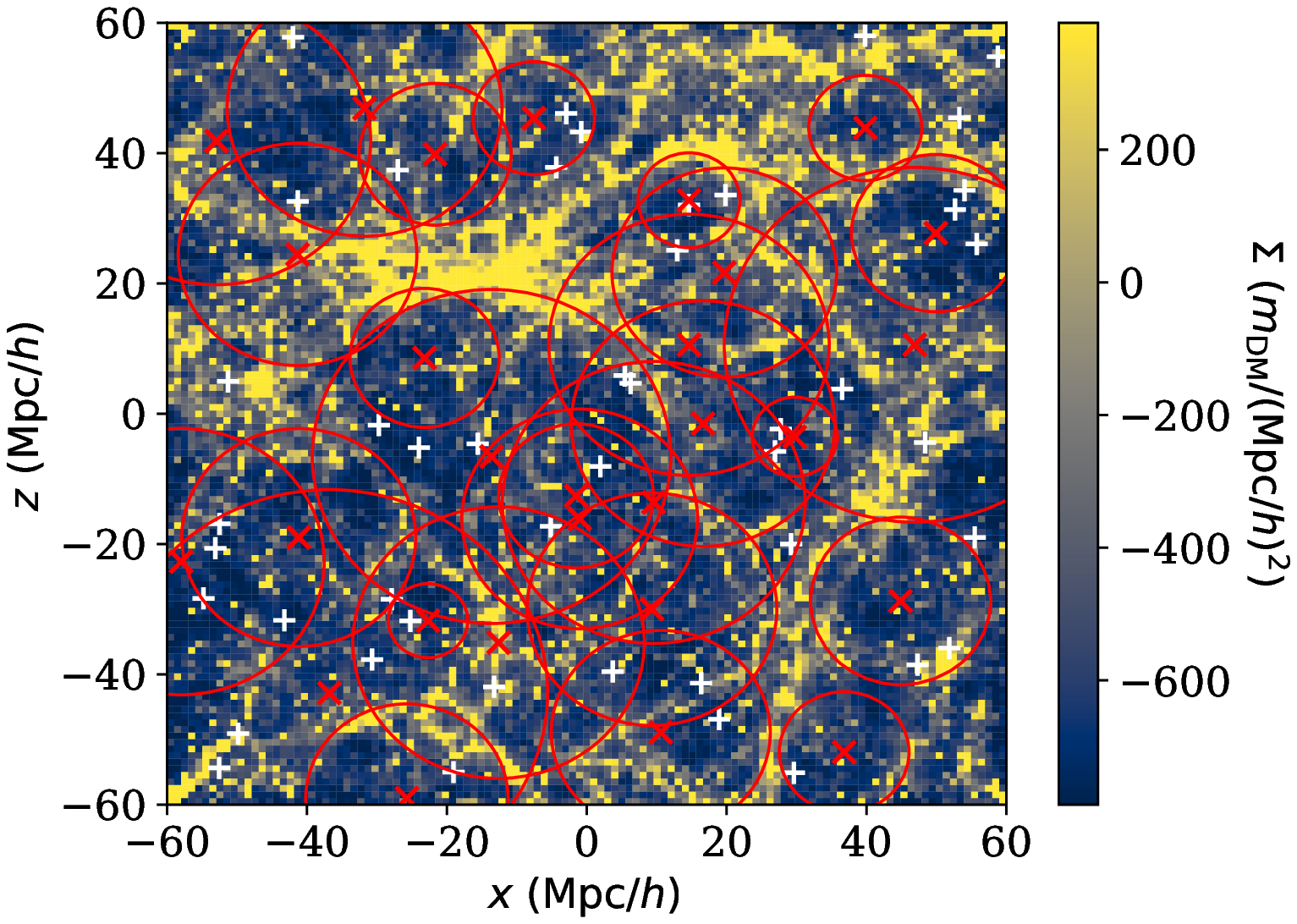}
  \includegraphics[width=0.95\columnwidth]{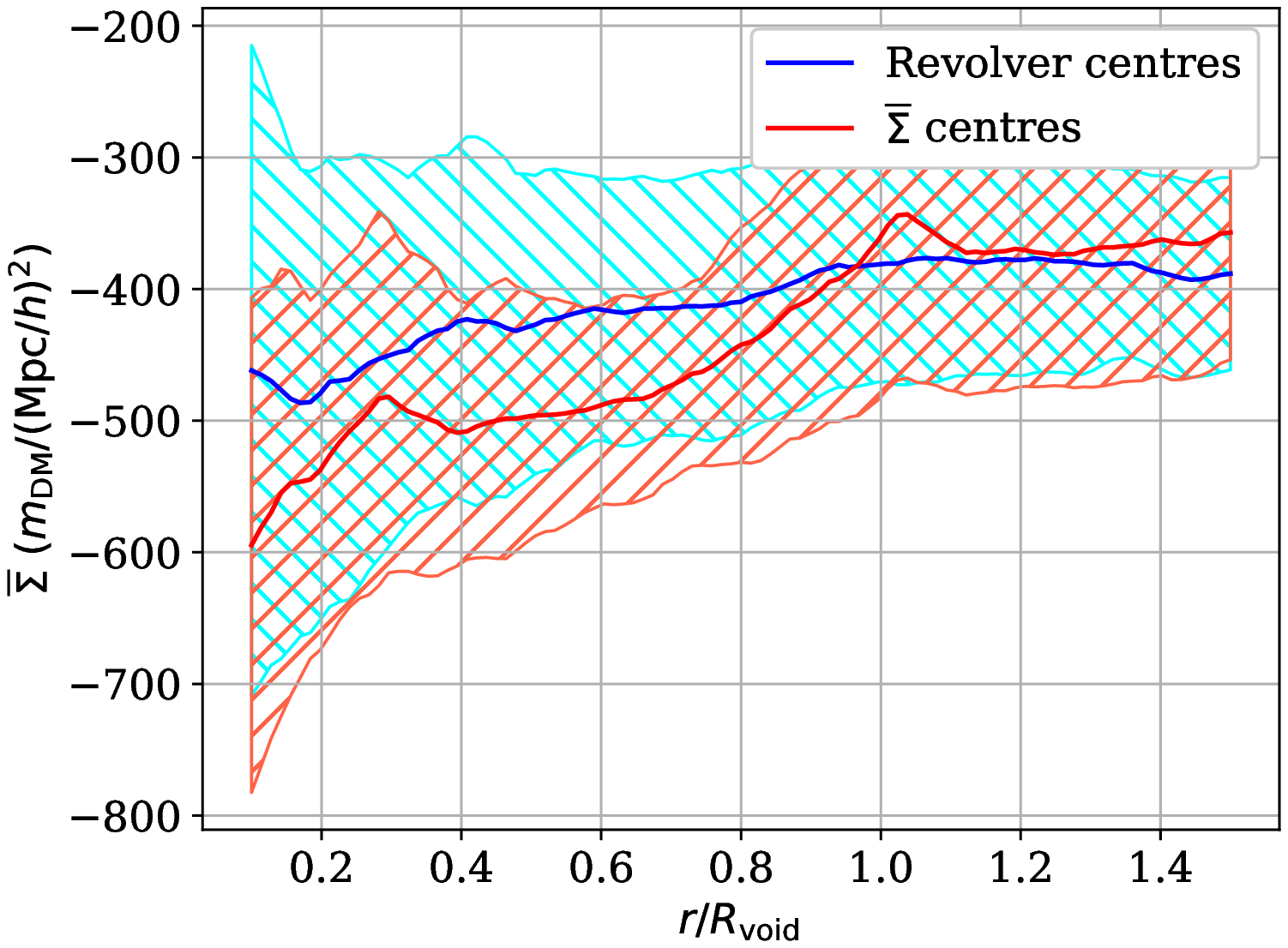}
  \caption{
    \emph{Upper panel:} surface overdensity $\Sigma$ projected along the line of sight.
    White $+$ symbols represent the $x,z$ centres of {\thrD} voids found with {\revolvername}.
    Red $\times$ symbols represent the centres of {\twD} voids found in the surface overdensity (\SSS\ref{s-meth-mass-def}); red circles represent the walls of these (circular, by definition) voids.
    Some of the {\thrD} void centres are projected close to one another in the sky plane; our algorithm is not designed to distinguish these as independent voids.
    \emph{Lower panel:} radial void profiles of the surface overdensity $\barSigma$ (Eq.~\protect\eqref{eq-circ-avg}), normalised to the estimated void radius and then averaged, using the set of all (projected) {\thrD} void centres and radii (mean: blue curve; standard deviation: green \enquote*{$\backslash\backslash$} hatching; \enquote*{Revolver centre}) or using the set of all {\twD} void centres and radii (mean: red curve; standard deviation: red \enquote*{//} hatching).
    \label{fig-mass-def}}
\end{figure}

\begin{figure}
  \centering
  \includegraphics[width=0.65\columnwidth]{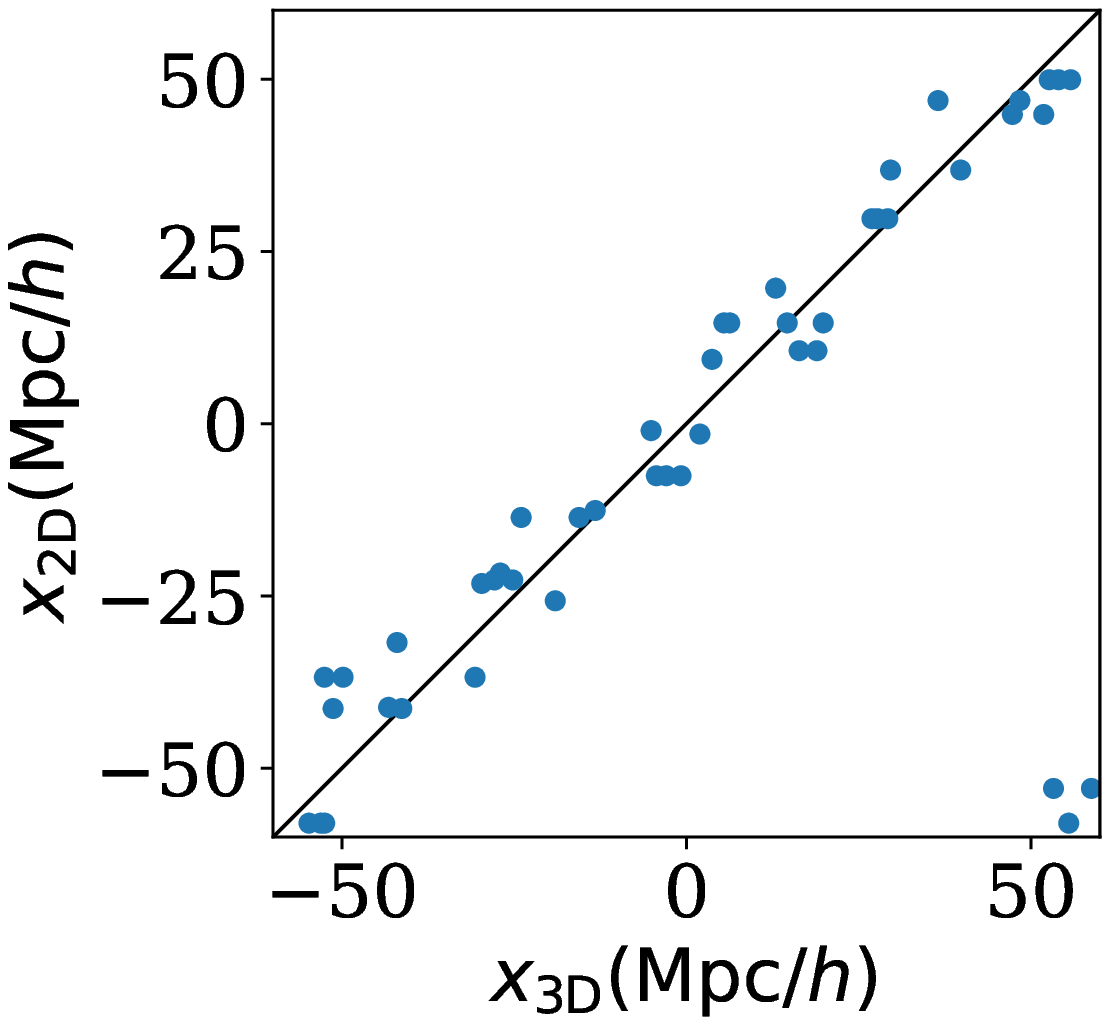}
  \includegraphics[width=0.65\columnwidth]{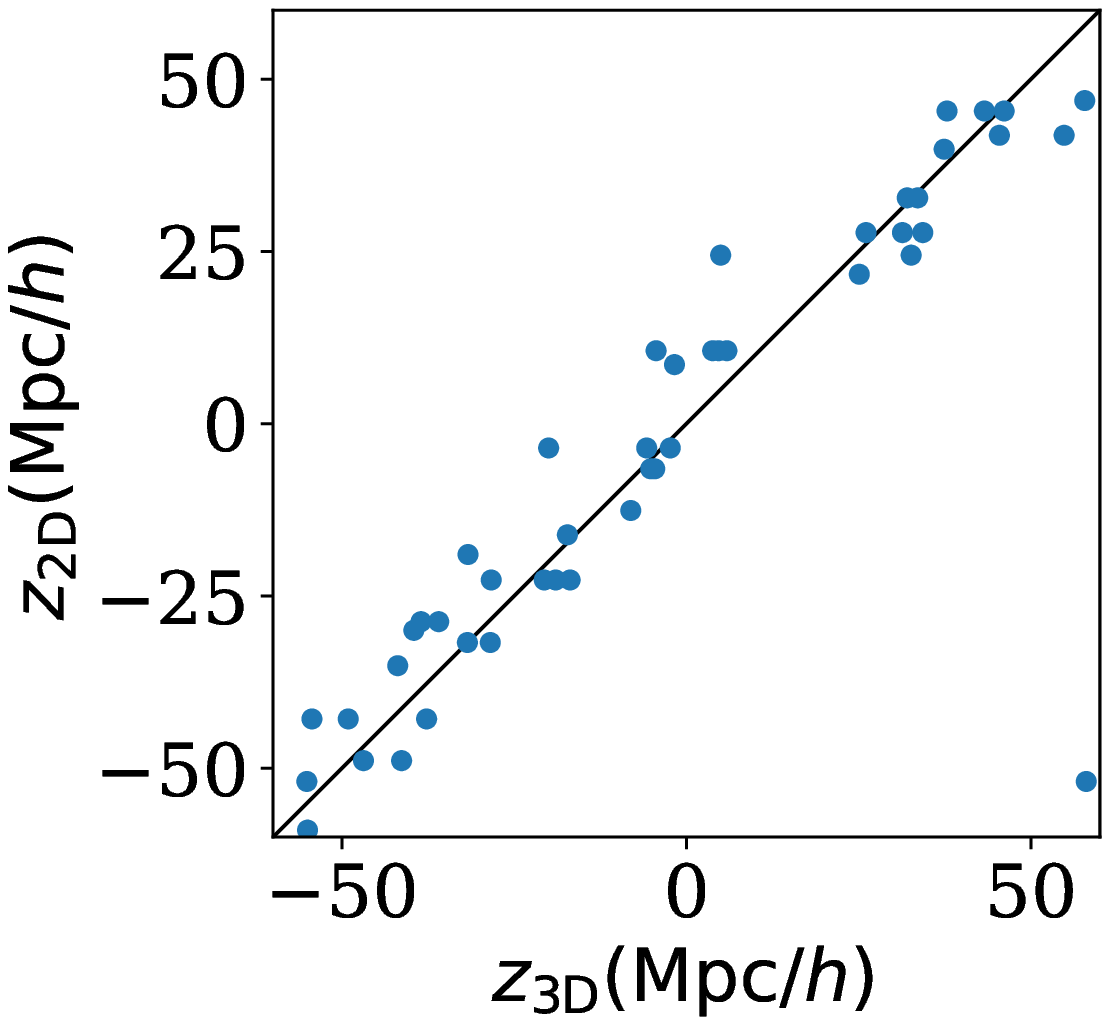}
  \includegraphics[width=0.65\columnwidth]{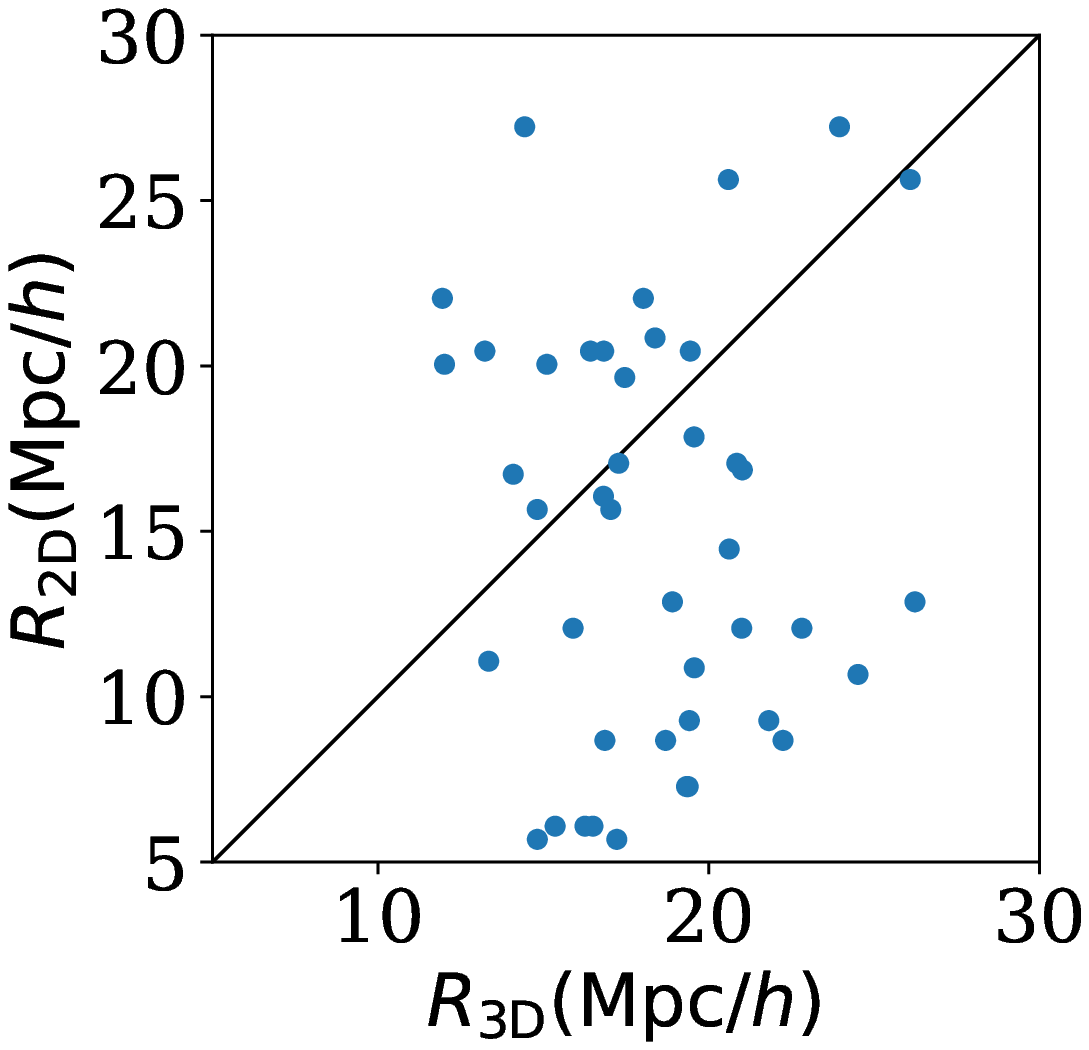}
  \caption{
    \emph{Top to bottom, respectively:}
    Given the a set of intrinsic {\thrD} voids in the galaxy distribution, sky-plane positions $x$ and $z$ and radii $R$ of the voids, and corresponding positions and radii of the {\twD} voids detected with the surface overdensity $\Sigma$ that best match these {\thrD} voids.
    The median $(x,z)$ $T^2$ distance for the best-matched voids, given a {\thrD} void (\SSS\ref{s-method-matches}), for detections with $\Sigma$ is $\XzMedianGivenThreeDUniqMassDefvalue$~Mpc/$h$.
    Equality is shown by a straight line in all three panels.
    The radii match poorly, with {\twD} radii mostly being less than the intrinsic {\thrD} radii.
    Plain-text data available at \href{\projectzenodofilesbase/void_matches_mass_def_given_3D.dat}{\projectzenodoid/void\_matches\_mass\_def\_given\_3D.dat}.
      \label{fig-bestmatch-mass-def}}
\end{figure}

Since voids in $N$-body simulations are characterised by small numbers of particles, the detection of individual voids, whether in the {\thrD} galaxy distribution or by a {\twD} detector in variables derived from the particle distribution, is in general numerically sensitive to small changes in machine arithmetic.
We performed a small number of independent full-pipeline simulations, retaining the same pseudo-random number seed, to investigate this qualitatively.
The re-simulated equivalent of the values listed in Table~\ref{t-match-significance} shows moderate variation with re-simulation on a given machine, and stronger variation between different machines.
We describe our results taking into account our small-scale estimates of their reproducibility, and use the word \enquote*{robust(ly)} to indicate cross-machine reproducibility.

We find $P^X_{x,z}(2D|3D)$ values (robustly) indicating significant match distributions in all four cases, with $P^\Sigma_{x,z}(2D|3D) < 0.001$, and $P^X_{x,z}(2D|3D) < 0.0001$ for $X \in \{\bargamma, \theta, \lvert\sigma\rvert \}$.
Thus, we find that given the {\thrD} voids found with the watershed algorithm in the galaxy distribution, the sky plane positions of the {\twD} voids found using the surface overdensity $\Sigma$ are significantly closer to the former than they would be if the same number of {\twD} void positions were chosen randomly.
In other words, we have a significant response to question (ii) for $\Sigma$.
This is reassuring, because it shows that despite the projection effects of multiple voids and their aspherical shapes, the centres of the intrinsic {\thrD} voids can be recovered in the {\twD} $\Sigma$ distribution.

Moreover, we find that for the weak-lensing tangential shear $\bargamma$, and for both the Sachs optical scalar expansion $\theta$ and shear $\lvert\sigma\rvert$, the centres of the {\twD} voids represent the {\thrD} void centres to high significance.
Thus, any of the four parameters should be usable to re-detect the void centres known from the {\thrD} voids.

In contrast, if we start with the {\twD} photometric map and predict the centres of the {\thrD} voids, we only find (Table~\ref{t-match-significance}) the weak-lensing tangential shear $\bargamma$, the Sachs expansion $\theta$ and the Sachs absolute shear $\lvert\sigma\rvert$ to significantly and robustly reveal underlying {\thrD} voids, with $P^X_{x,z}(3D|2D) \ll 0.01$ in all three cases.
Comparison with $P^\Sigma_{x,z}(3D|2D)$ in Table~\ref{t-match-significance}, for the surface overdensity, shows that discovering a {\thrD} void thanks to its {\twD} signature is less likely with $\Sigma$.
In other words, in answering question (i), use of our algorithm with any of the three geometrical optics parameters is more likely to reveal the sky-plane position of the {\thrD} void than using $\Sigma$.

These results show that the intrinsic {\thrD} void signal yields detectable void centres with our algorithm in not only the projected (\twD) surface overdensity $\Sigma$, inferrable from photometric maps with only a mass-to-light ratio assumption, but also in the {\twD} maps of weak-lensing and Sachs optical shear parameters.
If additional information, such as spectroscopic or photometric redshift information, is available, then combining that information with lensing analyses of the data should lead to tighter constraints on the (partly invisible) underdensity distributions, as opposed to using galaxies' sky positions and redshifts alone.

Moreover, in the absence of galaxy redshift information, {\twD} maps should yield constraints on the mass distribution, at least in the case of $\bargamma$ and $\theta$.
However, while the void centres are detected, the radii are poorly constrained from either {\thrD} or {\twD} maps.

We examine these results and caveats more closely in the following sections.

\subsection{Surface overdensity $\Sigma$} \label{s-results-mass-def}

The upper panel of Fig.~\ref{fig-mass-def} shows the map of the surface overdensity $\Sigma$, together with sky-plane centres of the intrinsic {\thrD} voids of the galaxy distribution and the {\twD} voids detected via $\Sigma$ as described in \SSS\ref{s-meth-mass-def}.
The correspondence between these, formalised in Table~\ref{t-match-significance}, can be inspected qualitatively by judging if a {\thrD} void centre (white $+$) has a {\twD} void centre (red $\times$) more close to it than a randomly placed point.
Of the $N_{3D} = \MassDefNVoidRevolvervalue$ intrinsic galaxy voids, only $N_{2D}^\Sigma = \MassDefNVoidvalue$ {\twD} voids are detected (Table~\ref{t-N-detections}).
The fact that $N_{2D}^\Sigma < N_{3D}$ is expected, since we did not design our algorithm to distinguish voids that are nearly concentric when projected to the sky plane.

\begin{figure}
  \includegraphics[width=0.95\columnwidth]{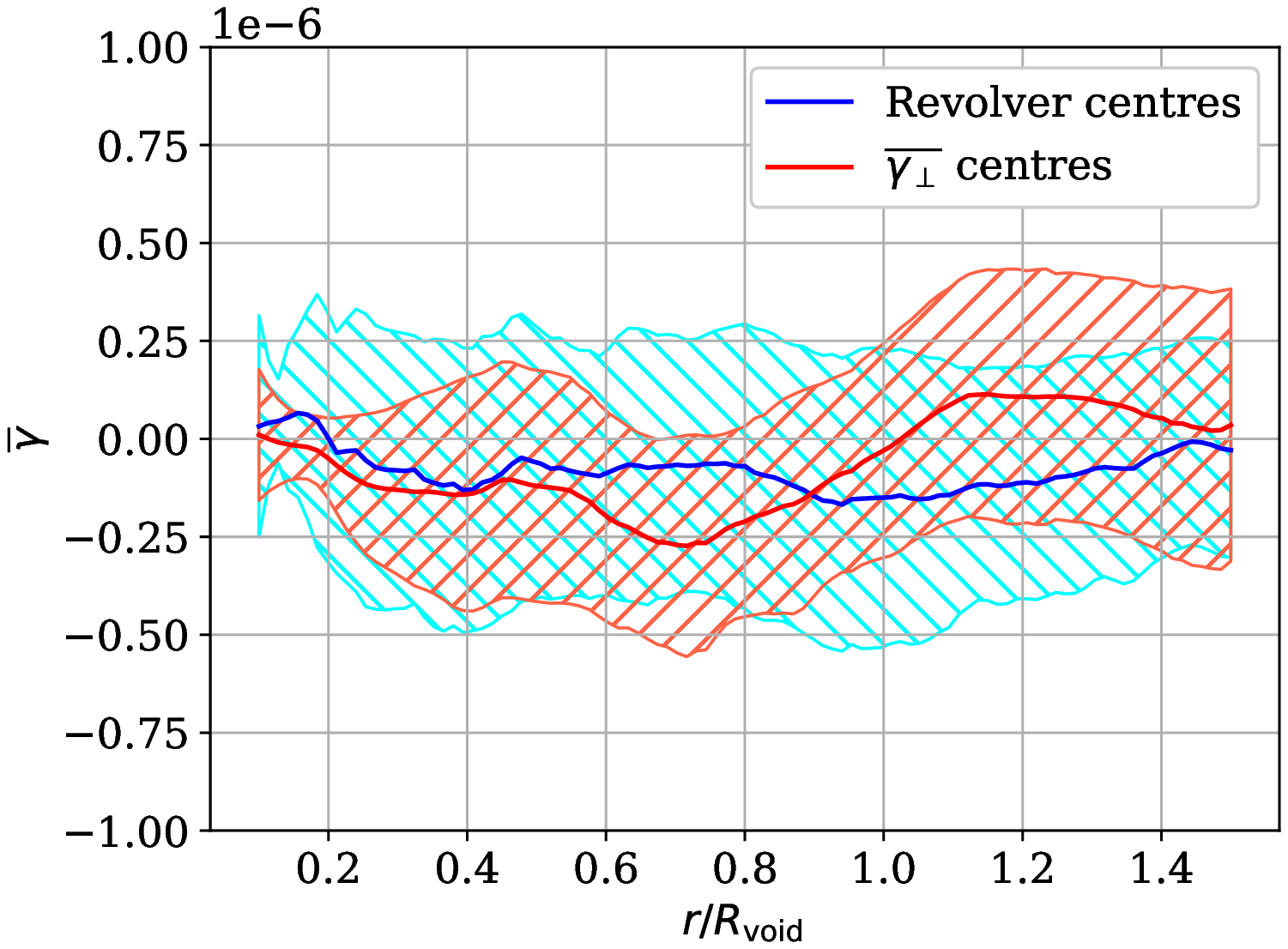}
\caption{
    Radial void profiles of the weak-lensing shear $\bargamma$, as in the lower panel of Fig.~\protect\ref{fig-mass-def}, for {\thrD} (projected) and {\twD} ($\bargamma$) sets of void centres.
    \enquote*{1e--6} indicates a factor of $10^{-6}$ in the vertical scale (and similarly in Figs~\protect\ref{fig-theta} and \protect\ref{fig-optical-scalars-mag-sigma} below).
    A map for $\bargamma$ is not shown, since the map of weak-lensing mean tangential shear $\bargamma(r,\hat{n})$ is redetermined for each possible void centre $\hat{n}$.
    \label{fig-weak-lens-gamma}}
\end{figure}

\begin{figure}
  \centering
  \includegraphics[width=0.65\columnwidth]{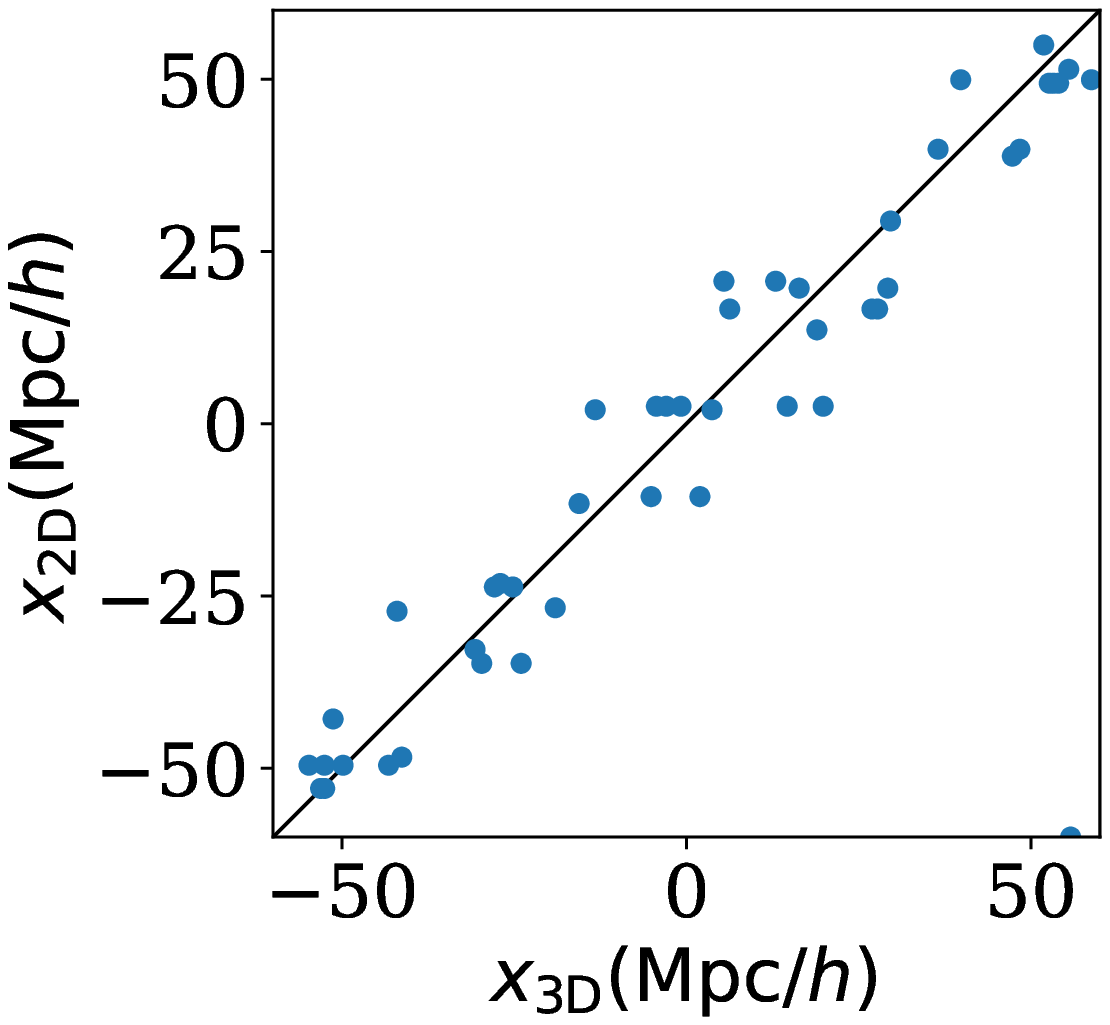}
  \includegraphics[width=0.65\columnwidth]{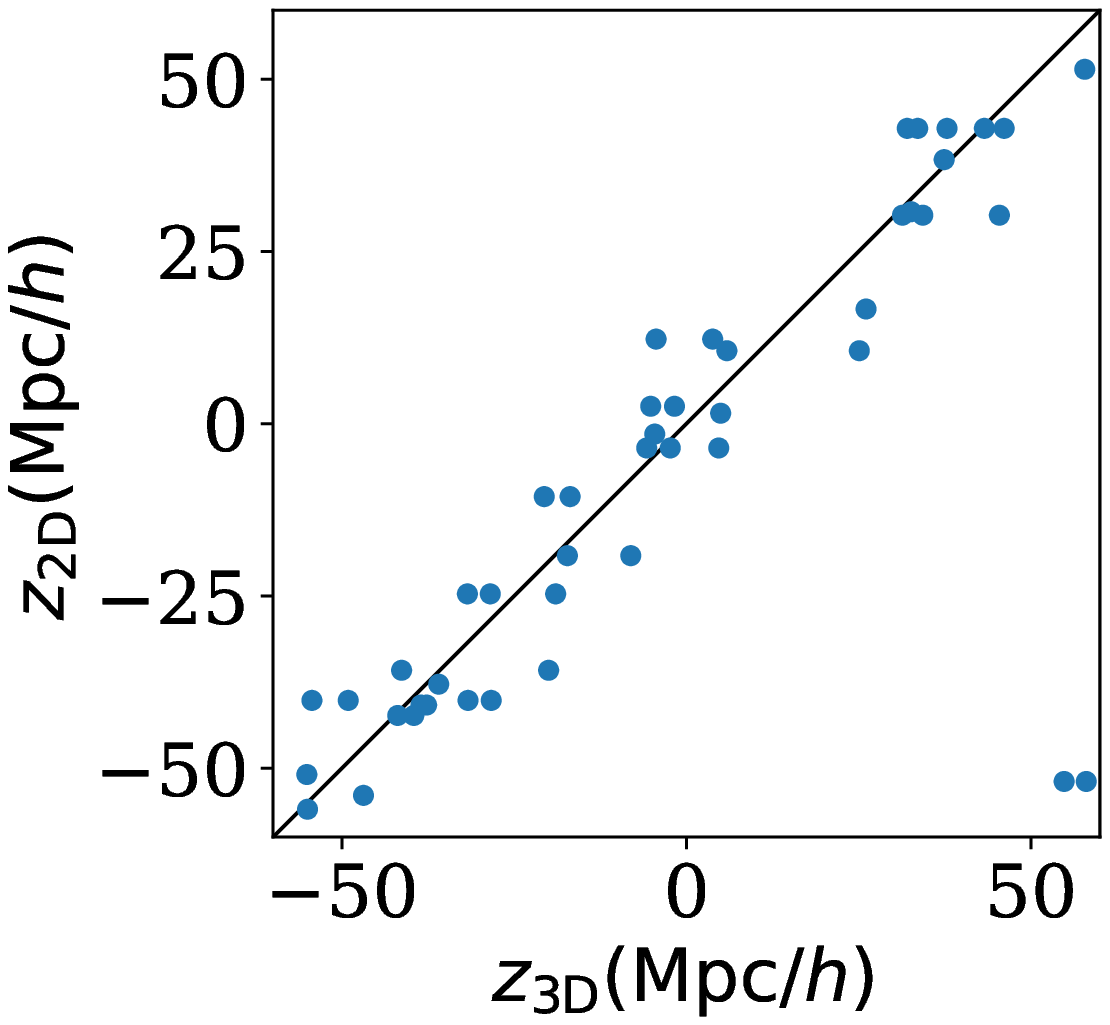}
  \includegraphics[width=0.65\columnwidth]{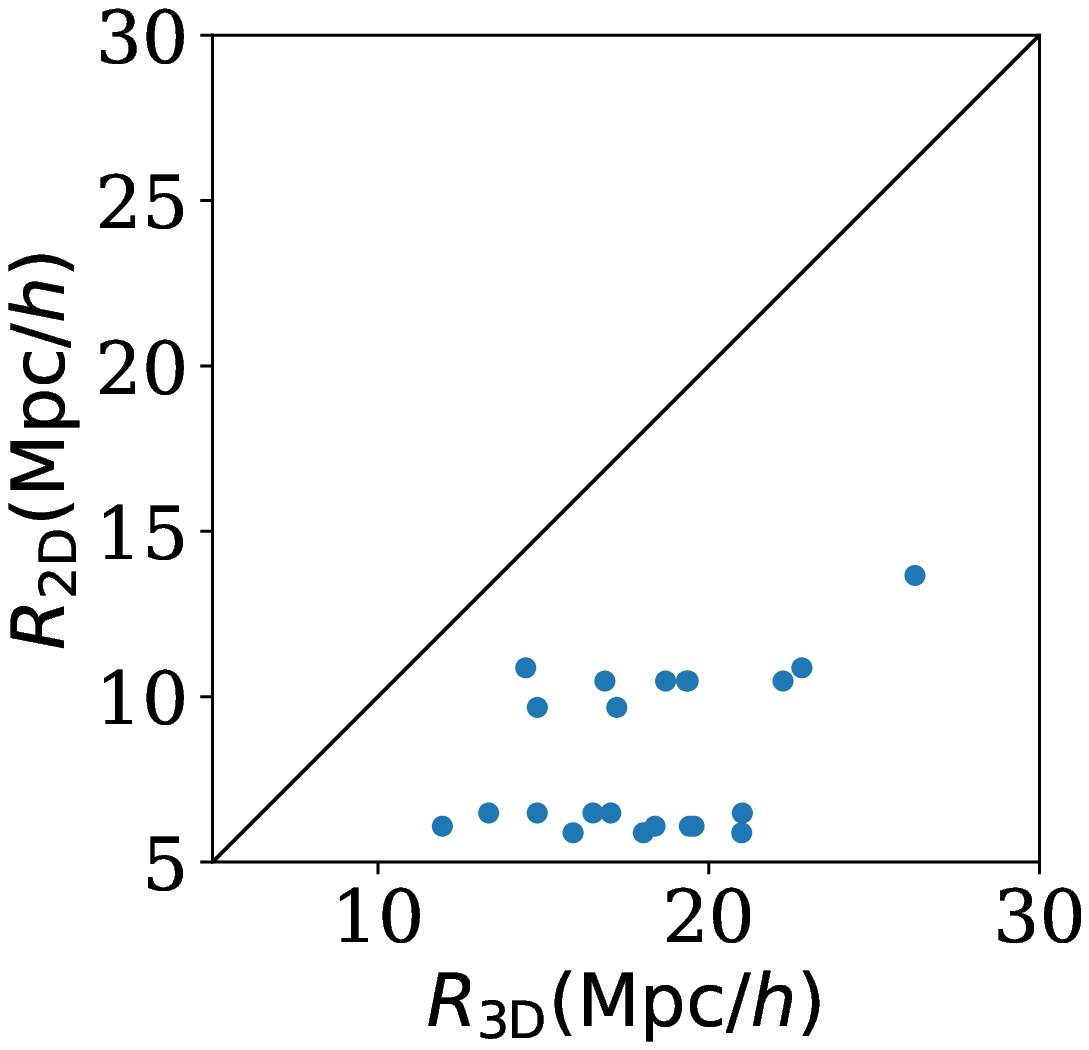}
  \caption{
    \emph{Top to bottom, respectively:}
    For each intrinsic {\thrD} void, sky-plane positions $x$ and $z$ and radii $R$ of the best-matched {\twD} void detected with the weak-lensing shear $\bargamma$, as in Fig.~\protect\ref{fig-bestmatch-mass-def}.
      The median $(x,z)$ $T^2$ distance for the best-matched voids, given a {\thrD} void (\SSS\ref{s-method-matches}), for detections with $\bargamma$ is $\XzMedianGivenThreeDUniqGammavalue$~Mpc/$h$.
      All {\twD} radii are lower than those of the {\thrD} voids that they correspond to.
      Plain-text data available at \href{\projectzenodofilesbase/void_matches_gamma_given_3D.dat}{\projectzenodoid/void\_matches\_gamma\_given\_3D.dat}.
    \label{fig-bestmatch-weak-lens}}
\end{figure}

The lower panel of Fig.~\ref{fig-mass-def} shows the $\barSigma$ profiles averaged over all the {\twD} centres, and, independently, averaged over all the {\thrD} centres (projected by ignoring the $y$ coordinate).
Comparison of these two curves (and their standard deviations, hatched) shows how well we might expect our algorithm to perform.
The profiles for the {\thrD} centres are those that would be detected if the algorithm were perfect in recovering the intrinsic voids, leaving aside the difference that the {\twD} detection uses the full dark matter particle distribution, while the {\thrD} detection is galaxy based.
It is clear that projection effects are significant: the mean profile (blue curve) does not show the sharp wall typical of voids.
It is also clear that we have found profiles in $\Sigma$ (red curve) that are stronger in contrast from minimum density to highest wall density than those of the intrinsic voids.
This suggests that improving the algorithm further based on the motivation of optimising a typical void-like profile, under the assumption of spherical shapes, would be unlikely to help further: strongly void-like profiles are already well detected.

The top two panels of Fig.~\ref{fig-bestmatch-mass-def} show the $x$ and $z$ coordinates (spanning the sky plane) of corresponding void centres, where the {\twD} void centres are those found to best match a given {\thrD} void, as described in \SSS\ref{s-method-matches}.
The existence of multiple {\thrD} voids whose best match is a single {\twD} void is clear in the diagram.
We interpret this as illustrating cases where {\thrD} voids are nearly aligned in projection, and thus detected as a single {\twD} void.

The bottom panel of Fig.~\ref{fig-bestmatch-mass-def} shows that void radii are very poorly recovered, and generally underestimated.
One factor is clearly the difficulty in distinguishing nearly concentric voids.
However, it is also likely that substructure is misidentified as void walls, leading to the underestimates.
Overall, the bottom panel of Fig.~\ref{fig-bestmatch-mass-def} shows that the radii of our intrinsic population of {\thrD} voids detected with {\revolvername} are reduced by about 5~Mpc/$h$, in an uncorrelated way with a big scatter, when redetected with $\Sigma$ as {\twD} voids.

\subsection{Weak-lensing shear $\bargamma$} \label{s-results-weak-lensing}

Using $\bargamma$, we find $N_{2D}^\gamma = \MassDefNVoidGammavalue$ {\twD} voids, i.e., roughly two thirds of the number of intrinsic galaxy voids, $N_{3D} = \MassDefNVoidRevolvervalue$ (Table~\ref{t-N-detections}).
Figure~\ref{fig-weak-lens-gamma} shows that the mean behaviour of a lensing profile in $\bargamma$ using the centres of the intrinsic {\thrD} voids is that $\bargamma$ starts near zero, decreases to negative values in the void and appears to (in the mean) reach a minimum at the wall radius found by {\revolvername}, before increasing to a maximum at a somewhat greater radius.
This is reasonable, given the definition fo $\bargamma$.
Figure~\ref{fig-weak-lens-gamma} shows that the {\twD} voids also have a (mean) $\bargamma$ profile that decreases and then increases to zero, but the increase to zero occurs at lower fractions of the void radius.

Together, these profiles could be interpreted to suggest that applying a systematic correction factor to increase the void radius found when $\tildeSigma(r,\hat{n}) = \barSigma(r,\hat{n})$ (see Eq.~\eqref{e-defn-DeltaSigma}) might yield radii that better match those of the {\thrD} voids.
The lowest panel of Fig.~\ref{fig-bestmatch-weak-lens} is qualitatively consistent with this suggestion, as it shows that the {\twD} voids that are best matched to the {\thrD} voids have radii that are all smaller than the {\thrD} void radii.
However, Table~\ref{t-match-significance} shows that correspondence between the radii is insignificant.
As in the case of $\Sigma$, the projection of nearly concentric intrinsic voids, as well as obscuring effects from more distant overlapping voids, make the use of a single scaling correction for radii poorly motivated, except as a crude statistical correction.

The two upper panels of Fig.~\ref{fig-bestmatch-weak-lens} show what is quantified in Table~\ref{t-match-significance}: the sky-plane positions are recovered non-randomly to high statistical significance.
Moreover, for the reverse question (Table~\ref{t-match-significance}), $P^\gamma_{x,z}(3D|2D) \ltapprox 0.001$ appears to be robust against re-calculation and machine error, so the use of weak-lensing shear -- on its own -- to infer the presence of intrinsic {\thrD} galaxy-traced voids appears to be promising.

\begin{figure}
  \centering
  \includegraphics[width=0.95\columnwidth]{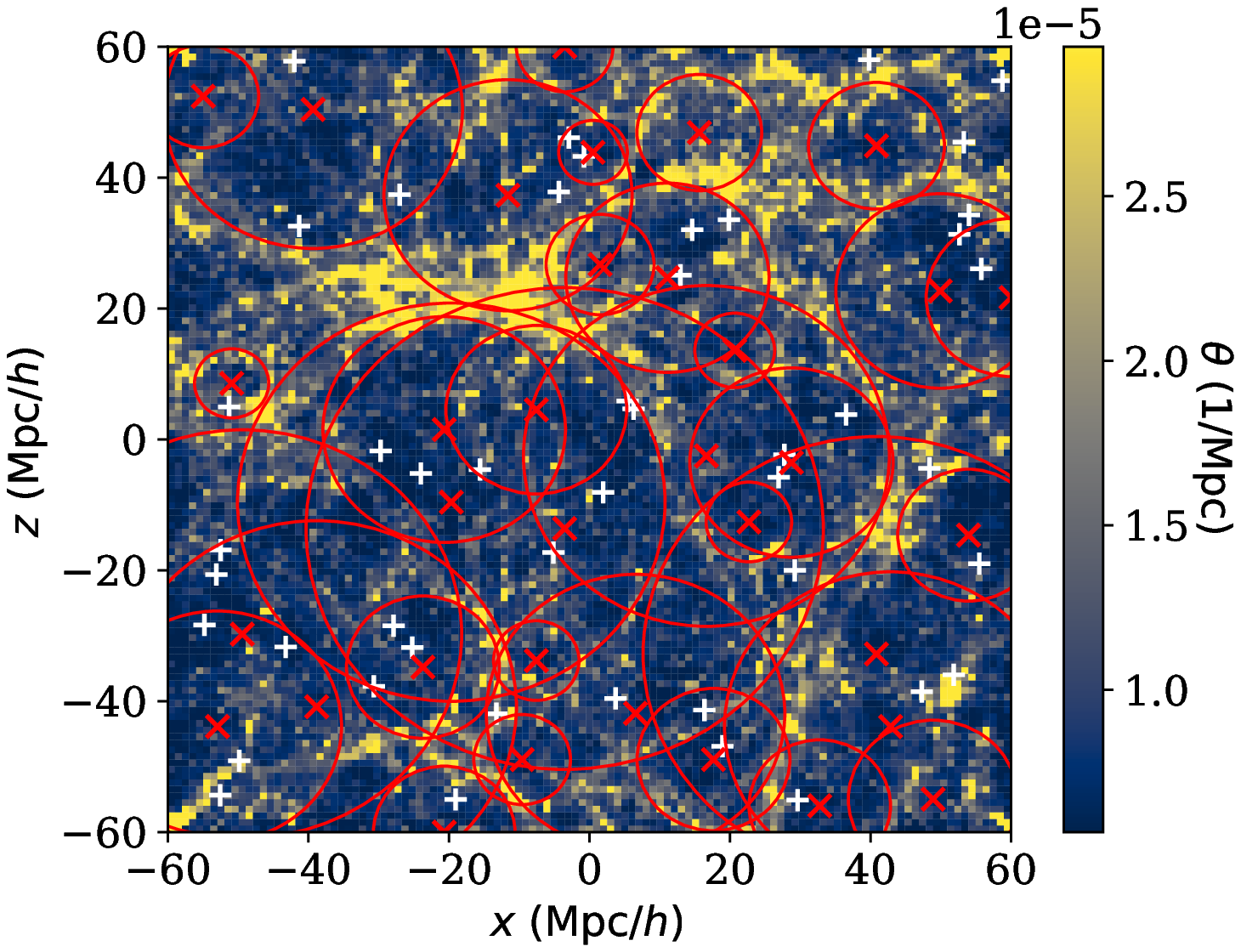}
  \includegraphics[width=0.95\columnwidth]{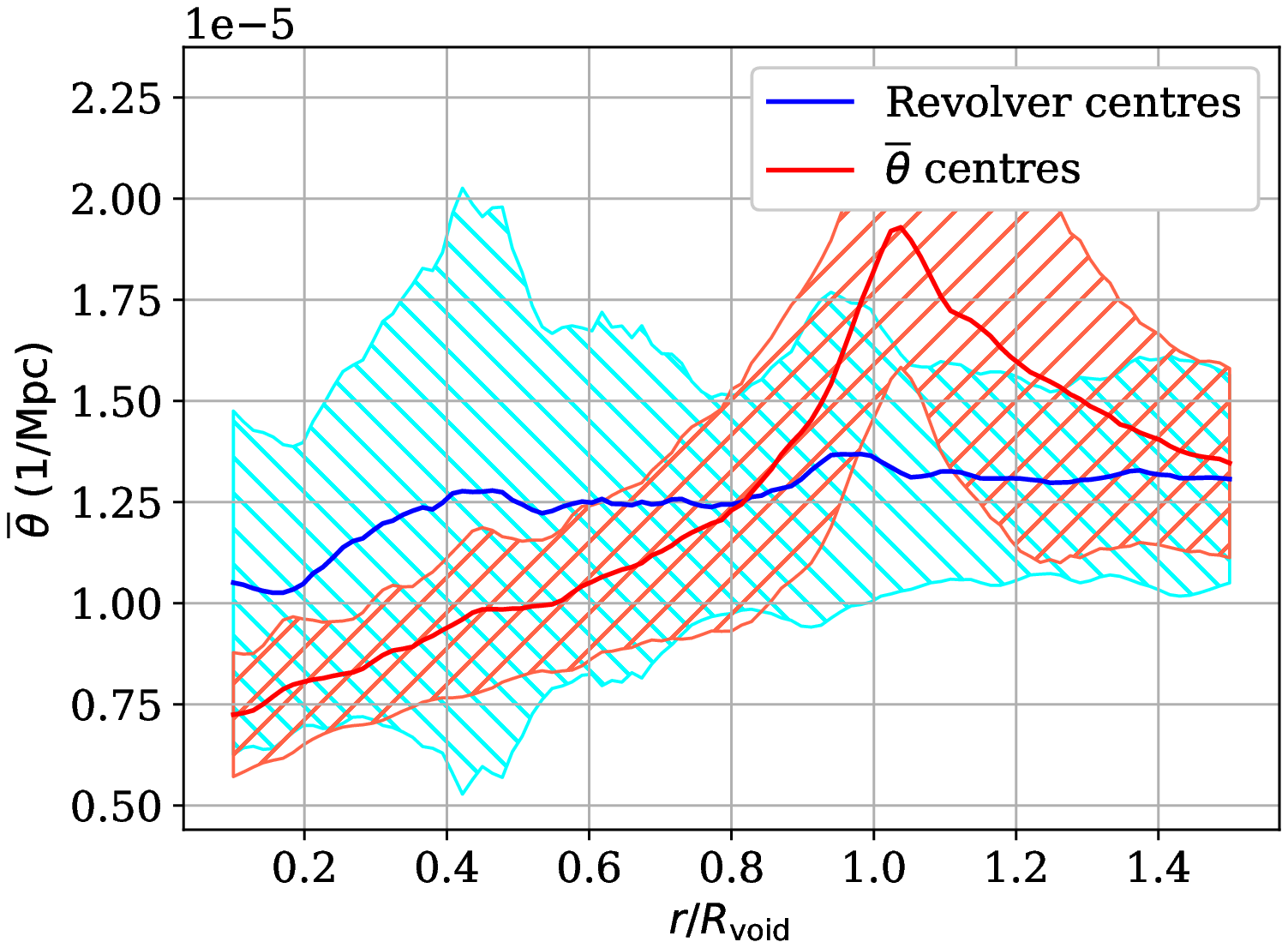}
  \caption{
    \emph{Upper panel:} Sachs expansion $\theta$,
    as for Fig.~\protect\ref{fig-mass-def},
    computed using Eqs~\eqref{eq-optical-scalars-theta} and \eqref{eq-optical-scalars-sigma},
    with
    white $+$ symbols for the $x,z$ centres of {\thrD} intrinsic galaxy voids and
    red $\times$ symbols for the centres of {\twD} voids detected with $\theta$.
    \emph{Lower panel:} Radial void profiles of $\theta$, as in the lower panel of Fig.~\protect\ref{fig-mass-def}, for {\thrD} ({\revolvername}) and {\twD} ($\theta$) sets of void centres.
    \label{fig-theta}}
\end{figure}

\begin{figure}
  \centering
  \includegraphics[width=0.65\columnwidth]{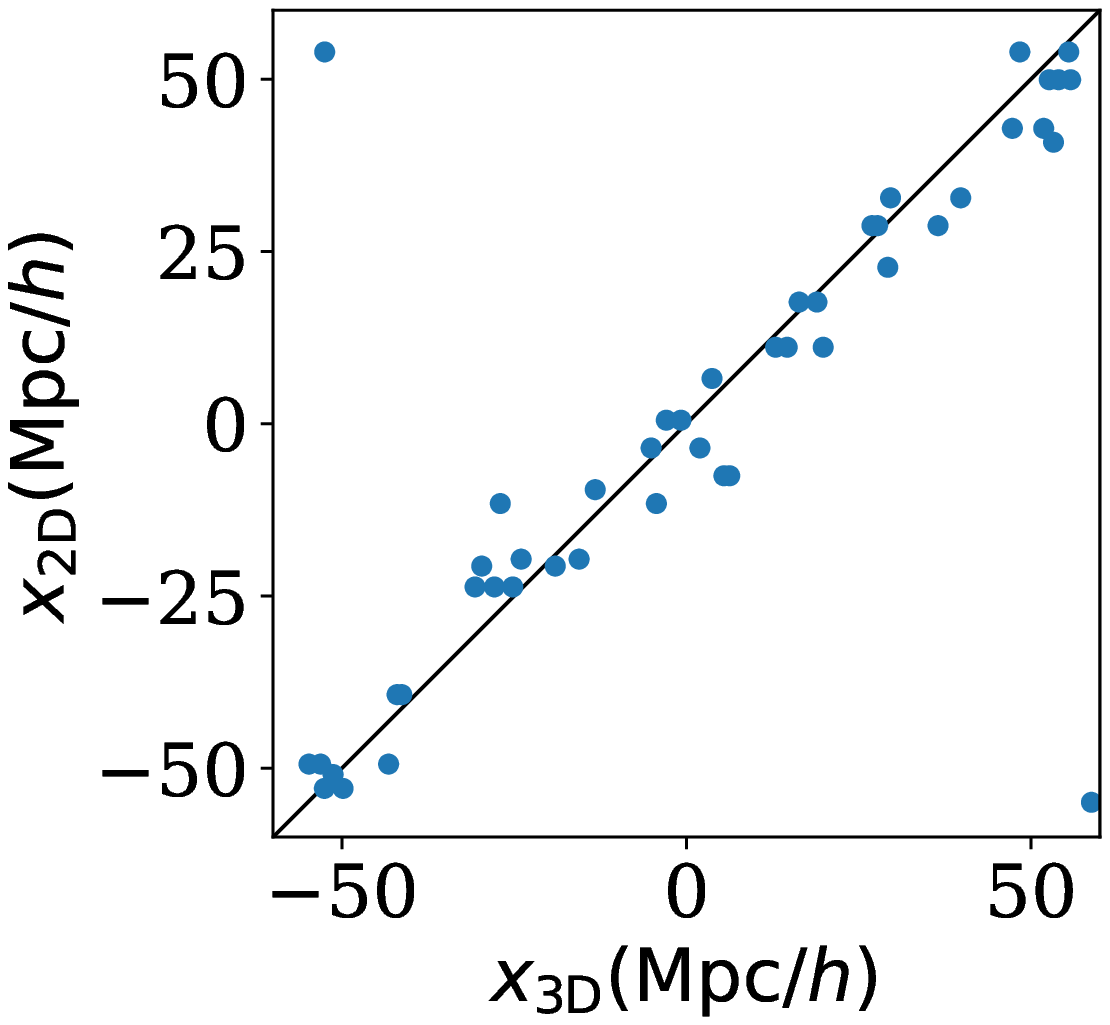}
  \includegraphics[width=0.65\columnwidth]{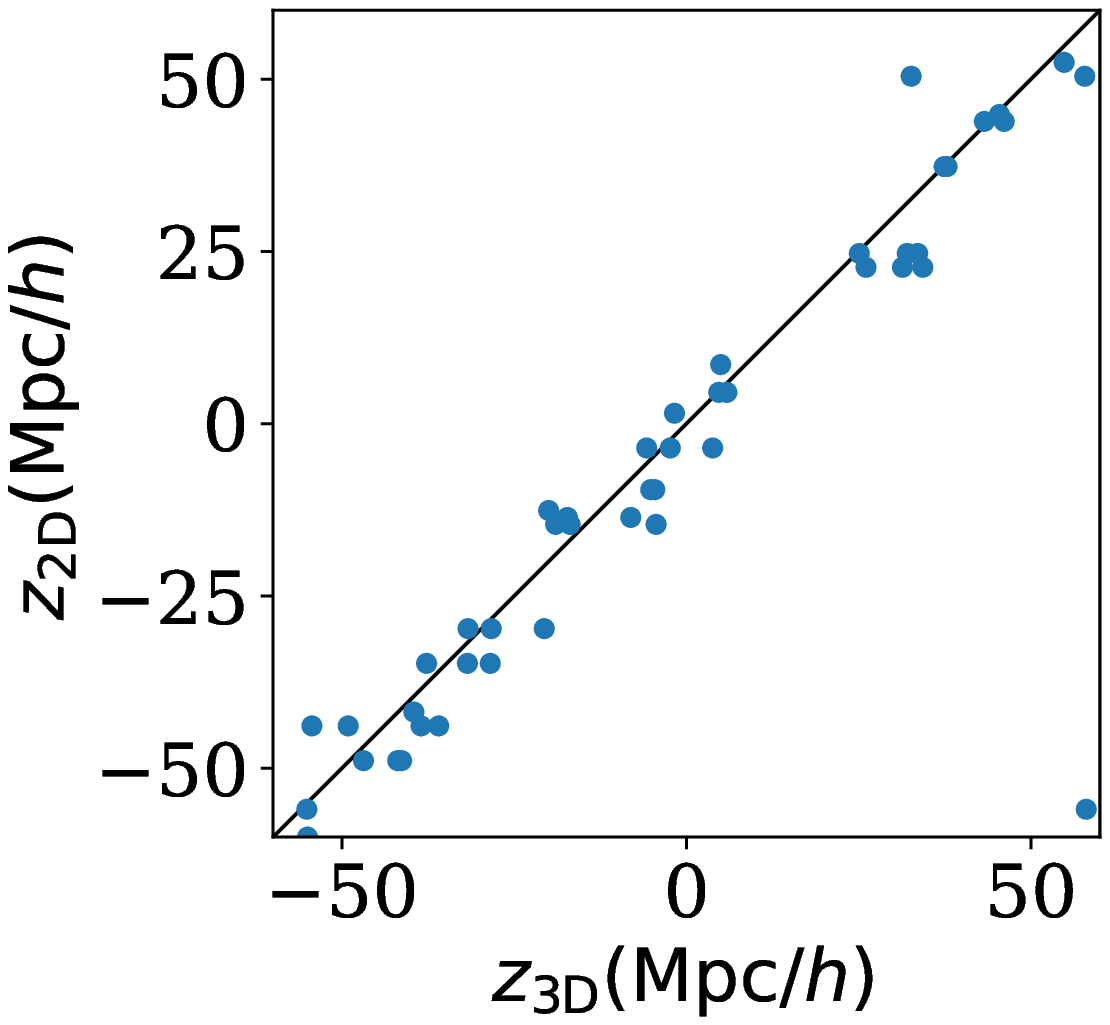}
  \includegraphics[width=0.65\columnwidth]{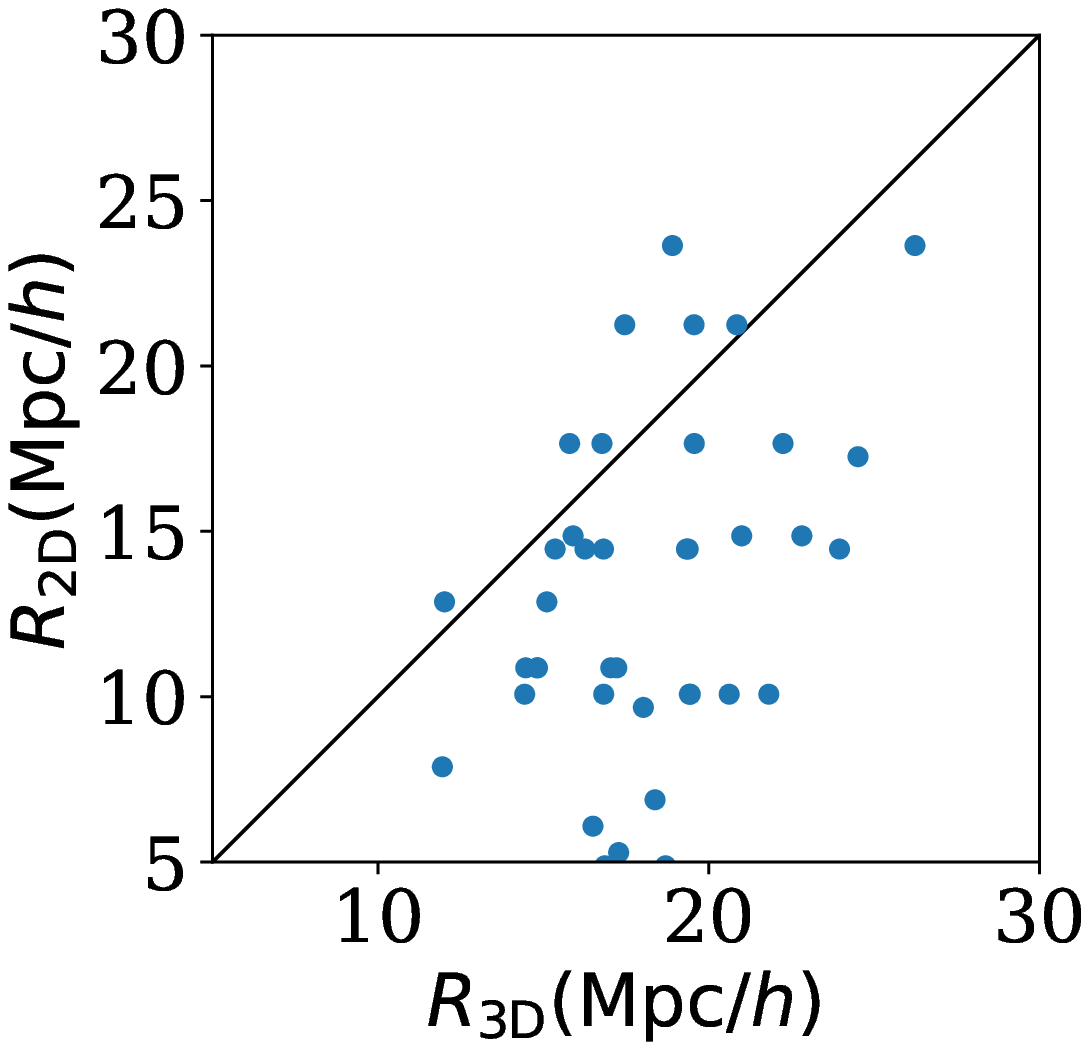}
  \caption{
    \emph{Top to bottom, respectively:}
    For each intrinsic {\thrD} void, sky-plane positions $x$ and $z$ and radii $R$ of the best-matched {\twD} void detected with the Sachs expansion $\theta$, as in Fig.~\protect\ref{fig-bestmatch-mass-def}.
    The median $(x,z)$ $T^2$ distance for the best-matched voids, given a {\thrD} void (\SSS\ref{s-method-matches}), for detections with $\theta$ is $\XzMedianGivenThreeDUniqExpvalue$~Mpc/$h$.
    The {\twD} radii have a much broader distribution than those of the intrinsic {\thrD} voids, with no obvious correlation.
    Plain-text data available at \href{\projectzenodofilesbase/void_matches_exp_given_3D.dat}{\projectzenodoid/void\_matches\_exp\_given\_3D.dat}.
      \label{fig-bestmatch-expansion}}
\end{figure}

\subsection{Optical scalars $\theta$ and $\lvert\sigma\rvert$} \label{s-results-opt-scal}

\subsubsection{Expansion $\theta$}
Figure~\ref{fig-theta} shows a map of the Sachs expansion $\theta$ and sky-plane centres of both the intrinsic voids and those detected via $\theta$.
As indicated in Table~\ref{t-match-significance}, given the {\thrD} voids, the best-matched {\twD} ($\theta$) voids are recovered to high significance via their sky-plane centres.
The top two panels of Fig.~\ref{fig-bestmatch-expansion} show the sky-plane matches.

However, we only find $N_{2D}^\theta = \OSNVoidExpvalue$ voids using $\theta$, many less than the intrinsic voids.
As with $\Sigma$ and $\bargamma$, a likely interpretation is projected concentricity of several voids and obscuration by other cosmic web structure.
The lower panel of Fig.~\ref{fig-theta} can be interpreted consistently with this hypothesis: the mean $\theta$ profile of the full set of intrinsic voids detected with {\revolvername} is very weak, which would be consistent with both effects.
The profile for {\twD} voids detected with $\theta$ is very strong, qualitatively resembling a typical void density profile, with a sharp (mean) wall.

The lowest panel of Fig.~\ref{fig-bestmatch-expansion} shows that the radii are again poorly correlated.
Again, this is consistent with the detections using $\Sigma$ and $\bargamma$, with the difference that the radii estimated with $\theta$ expand greatly from the instrinic voids' range of around 15--25~Mpc/$h$ to around 5--30~Mpc/$h$.
While to some degree these disagreements are likely to be induced by the problems of projection, it might also be possible that radii that are gravitationally realistic in terms of the potential $\Phi$ differ significantly from those traced by the {\thrD} galaxy distribution.
This is a question open for further study.

\begin{figure}
  \centering
  \includegraphics[width=0.95\columnwidth]{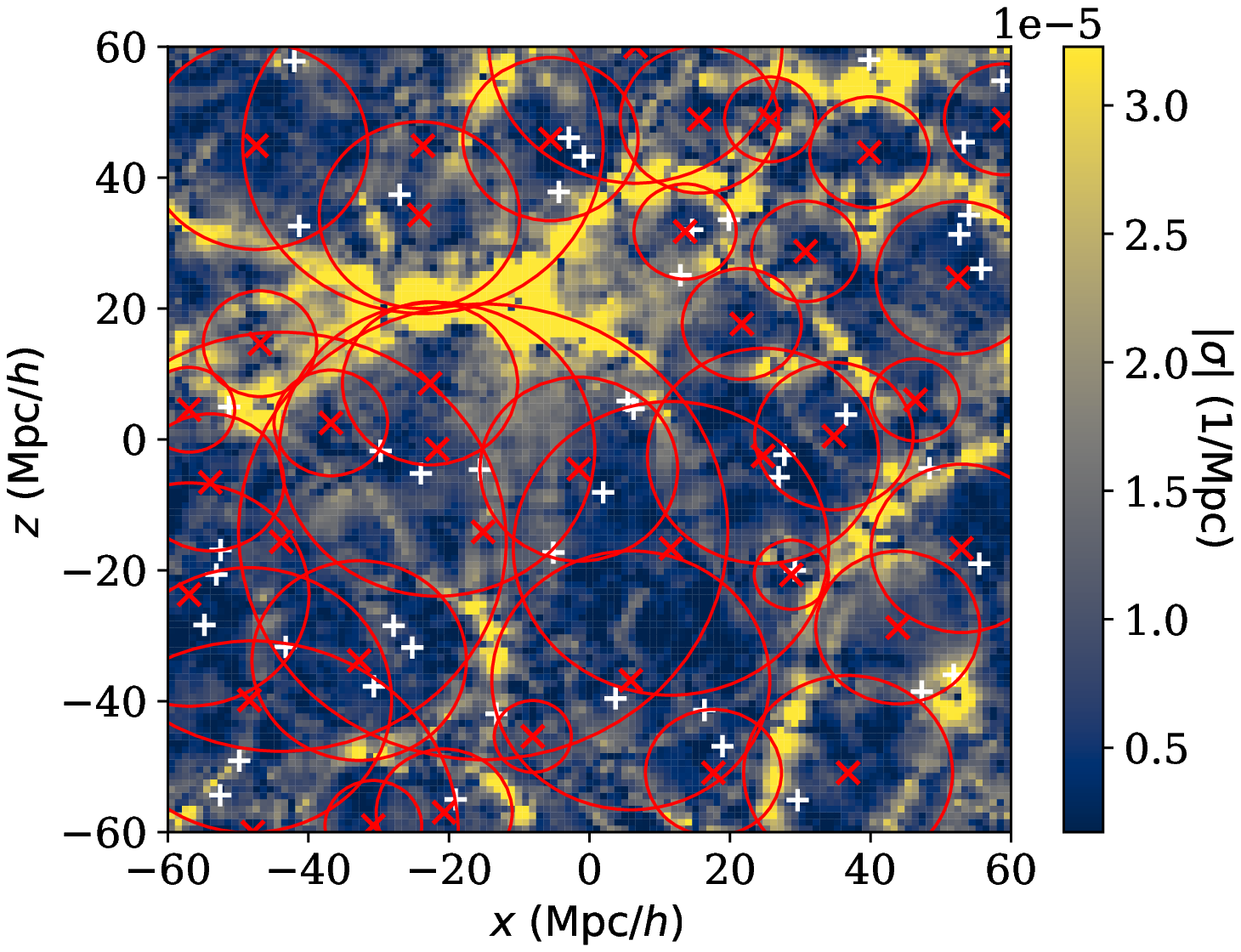}
  \includegraphics[width=0.95\columnwidth]{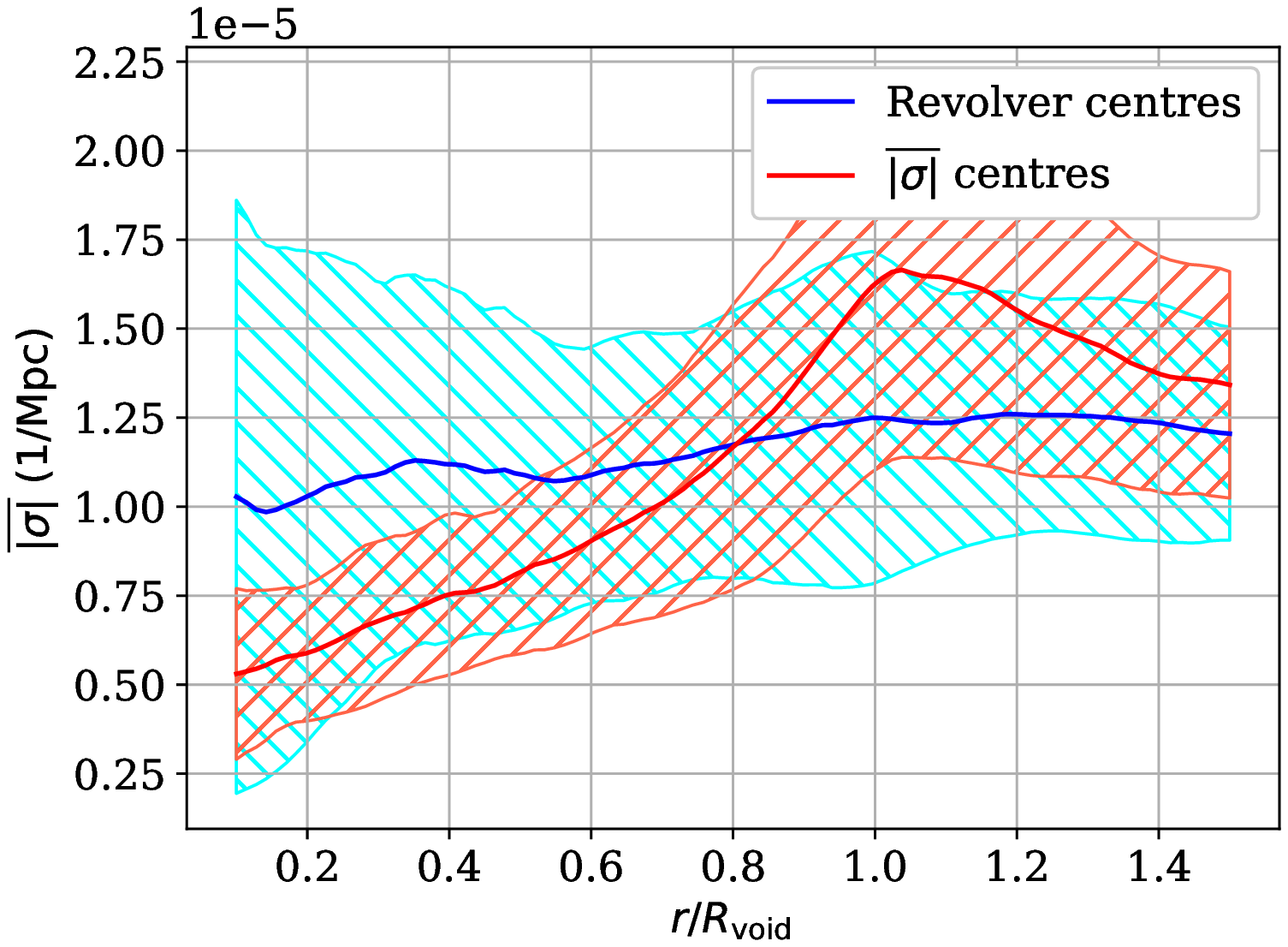}
  \caption{
    \emph{Upper panel:} Sachs shear $\lvert\sigma\rvert$,
    as for Fig.~\protect\ref{fig-mass-def},
    computed using Eqs~\eqref{eq-optical-scalars-theta} and \eqref{eq-optical-scalars-sigma},
    with
    white $+$ symbols for the $x,z$ centres of {\thrD} intrinsic galaxy voids and
    red $\times$ symbols for the centres of {\twD} voids detected with $\lvert\sigma\rvert$.
    \emph{Lower panel:} Radial void profiles of $\lvert\sigma\rvert$, as in the lower panel of Fig.~\protect\ref{fig-mass-def}, for {\thrD} ({\revolvername}) and {\twD} ($\lvert\sigma\rvert$) sets of void centres.
    \label{fig-optical-scalars-mag-sigma}}
\end{figure}

\begin{figure}
  \centering
  \includegraphics[width=0.65\columnwidth]{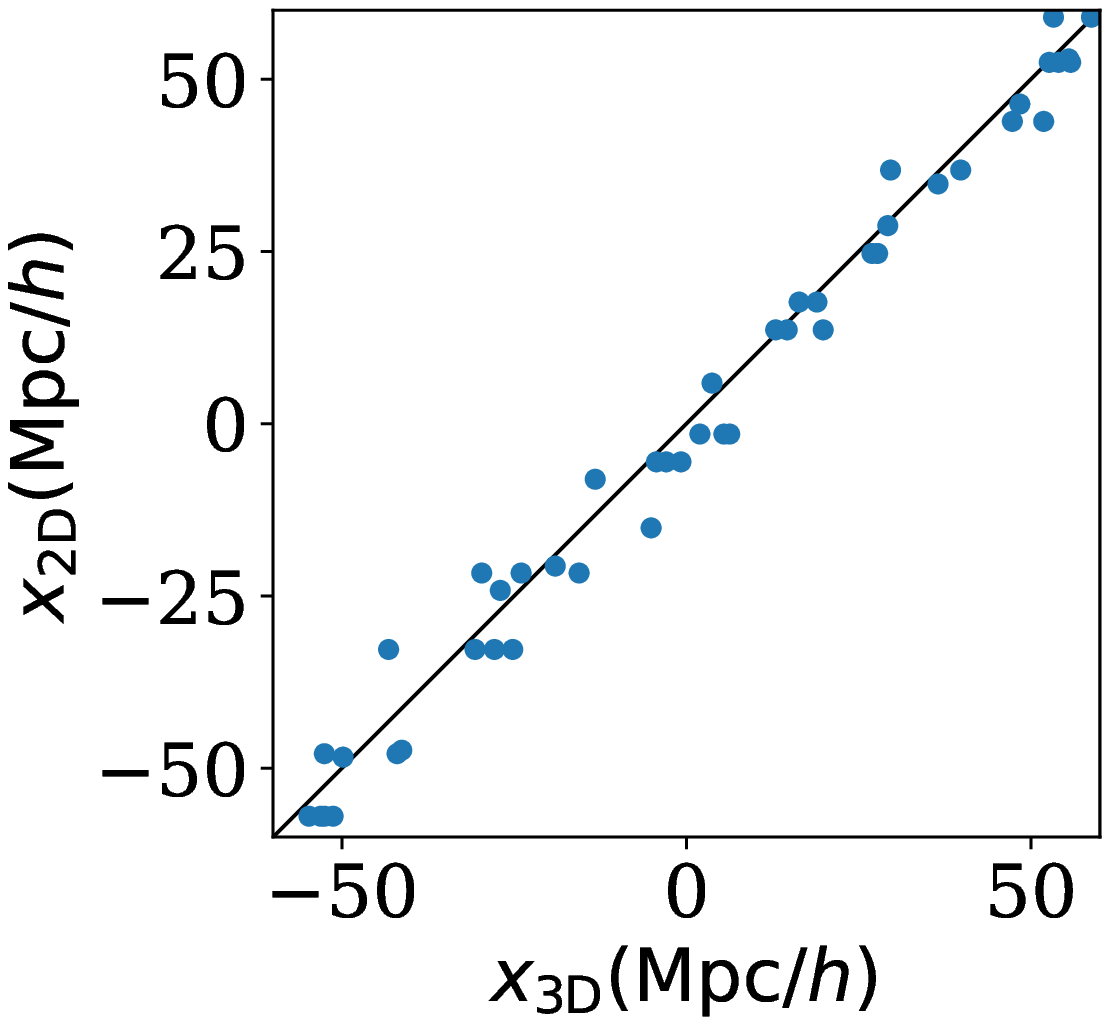}
  \includegraphics[width=0.65\columnwidth]{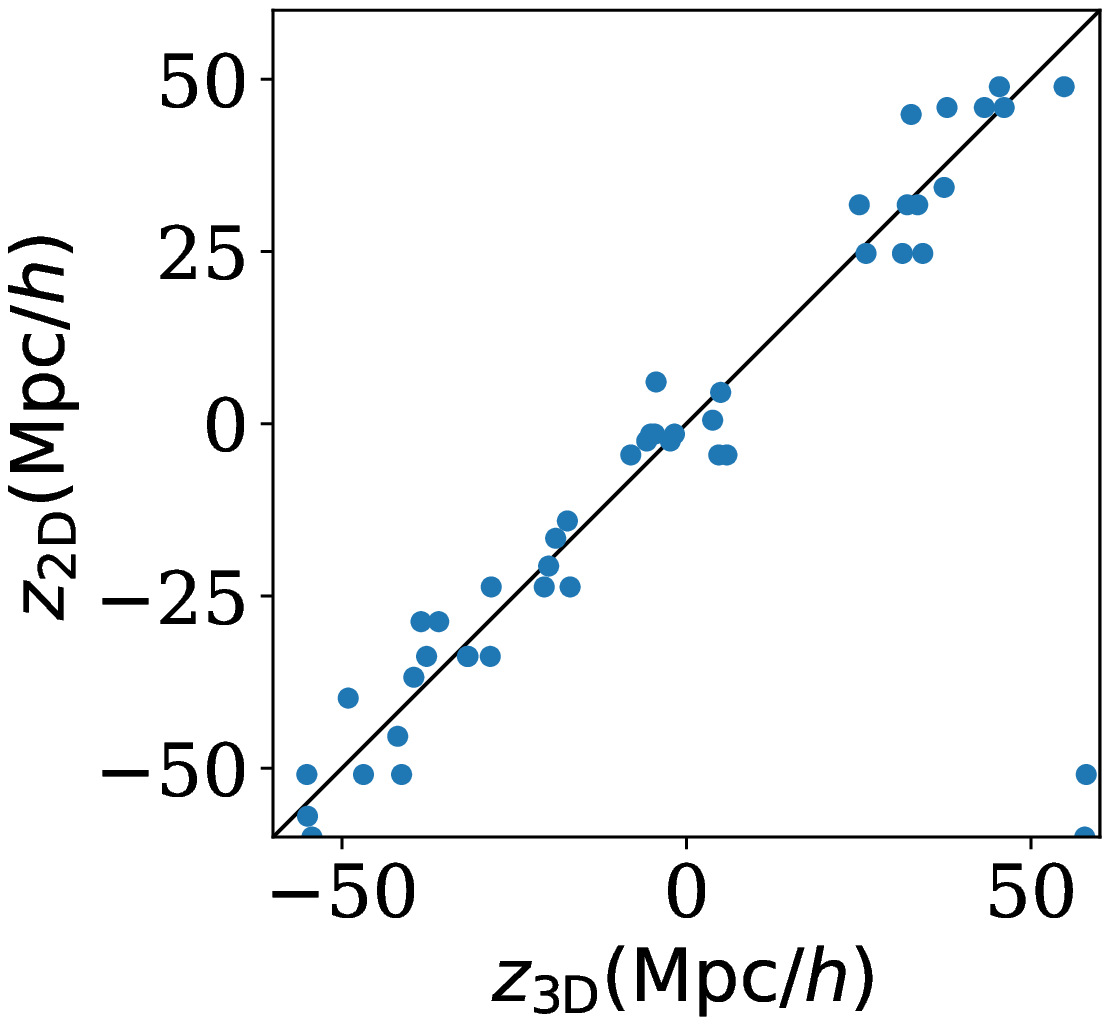}
  \includegraphics[width=0.65\columnwidth]{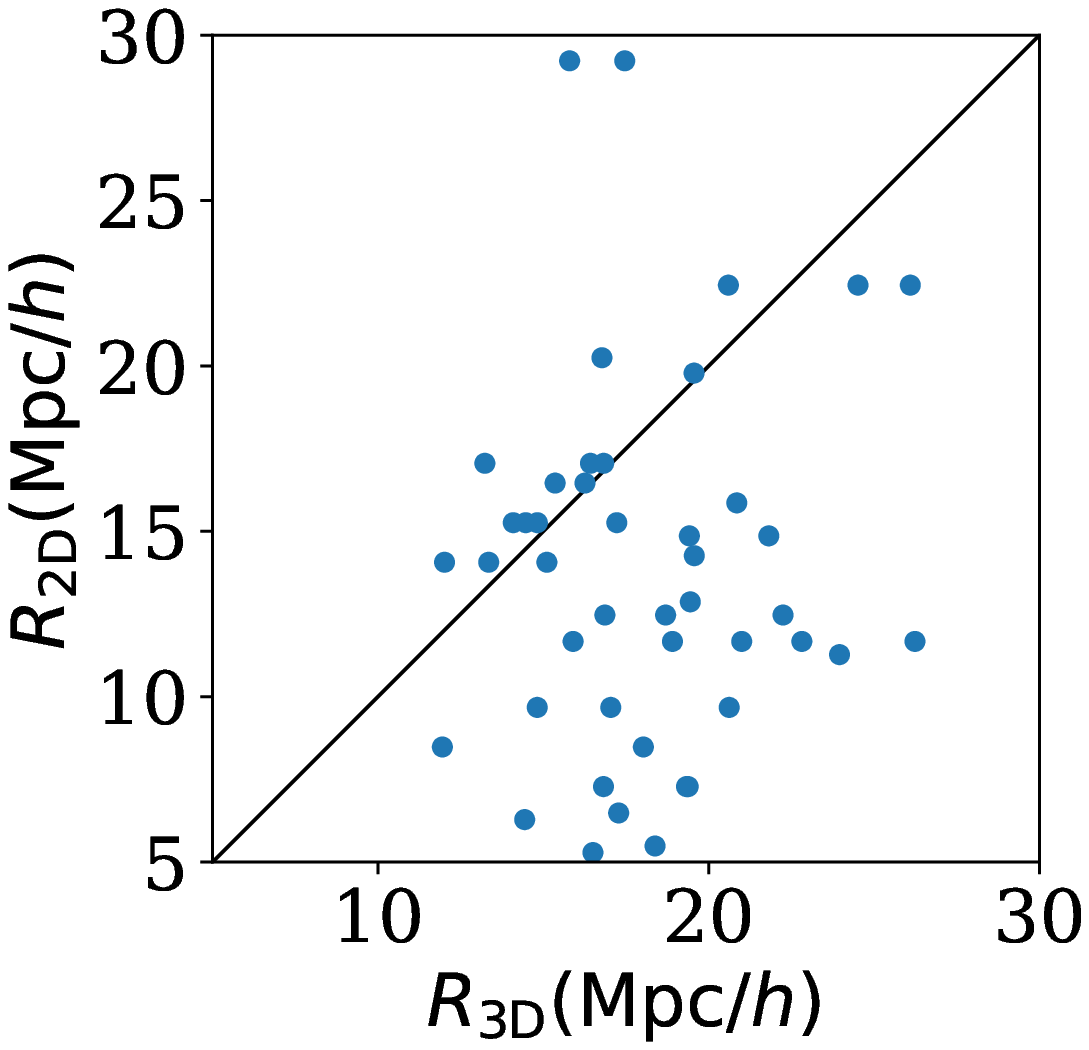}
  \caption{
    \emph{Top to bottom, respectively:}
    For each intrinsic {\thrD} void, sky-plane positions $x$ and $z$ and radii $R$ of the best-matched {\twD} void detected with the Sachs shear modulus $\lvert\sigma\rvert$, as in Fig.~\protect\ref{fig-bestmatch-mass-def}.
      The median $(x,z)$ $T^2$ distance for the best-matched voids, given a {\thrD} void (\SSS\protect\ref{s-method-matches}), for detections with $\lvert\sigma\rvert$ is $\XzMedianGivenThreeDUniqSigvalue$~Mpc/$h$.
      The {\twD} radii again have a broad distribution, as in Fig.~\protect\ref{fig-bestmatch-expansion}.
      Plain-text data: \href{\projectzenodofilesbase/void_matches_sig_given_3D.dat}{\projectzenodoid/void\_matches\_sig\_given\_3D.dat}.
      \label{fig-bestmatch-sigma}}
\end{figure}

\subsubsection{Sachs shear $\lvert\sigma\rvert$}

The upper panel of Fig.~\ref{fig-optical-scalars-mag-sigma} shows a map of the modulus of the Sachs shear, $\lvert\sigma\rvert$.
Again, Table~\ref{t-match-significance} shows that given an intrinsic void, the Sachs shear detects the voids' positions to high significance using our algorithm.
The lower panel of Fig.~\ref{fig-optical-scalars-mag-sigma} shows a qualitatively similar result to the use of the expansion $\theta$, in the sense that shear profiles for the full set of {\thrD} voids have a weak mean profile, while those for the voids detected in the {\twD} map of $\lvert\sigma\rvert$ show a strong void-like profile.
Taking into account the good sky-plane matches and poor radial matches shown in Fig.~\ref{fig-bestmatch-sigma}, a consistent interpretation is again that the {\twD} detected profiles are those that bypass both general obscuration and the confusion induced by voids that are nearly concentric in projection.

\section{Discussion} \label{s-discuss}

\subsection{Void lensing studies when intrinsic voids are known}
With the simulation presented here, we have shown that if intrinsic {\thrD} voids are known, then the effects of geometric-optics parameters should be detectable in the sky plane, enabling the study of the role that gravitational lensing plays in the voids.
In other words, we have shown a relation between voids in {\thrD} comoving space with their imprints left on maps of the projected and ray-traced variables.
Moreover, the lensing patterns induced by a void should provide feedback to better constrain the model of the void itself.
As argued by \cite{SanchezTwoDVoidFinding} using $\Sigma$ (and photometric redshifts to statistically limit the radial depth of the projection), this would confirm that a galaxy-traced void is a genuine underdensity of the dark matter density field.
Other weak gravitational lensing detectors, such as the Lyman~$\alpha$ forest \citep{Croft18LyAlphaLensing}, could also be compared to foreground galaxy-traced voids to check for consistency.

\subsection{Blind (redshift-free) searches for voids}
Without knowledge of spectroscopic or photometric galaxy redshifts, we currently can justify use of the azimuthally averaged tangential weak-lensing shear $\bargamma$ and of the Sachs expansion $\theta$ or shear $\lvert\sigma\rvert$ for analysis of a photometric survey with the intention of inferring the presence of {\thrD} voids, since all three of these robustly yield $P^X_{x,z}(3D|2D) \ll 0.01$ (Table~\ref{t-match-significance}).
Our calculation would appear to be the first time that the use of Sachs expansion maps has been shown to have the ability to reveal underlying voids.
\citet{Jeffrey2021DES3WeakLensing} studied the combined use of the usual weak-lensing convergence and shear in Dark Energy Survey (DES) photometry over 4143~deg$^2$, which appears to be equivalent to using $\Sigma$ and $\bargamma$, to reveal underlying voids, but did not appear to study the Sachs expansion.

Our algorithms can very likely be improved further.
For example, combining all four parameters, $\Sigma$, $\bargamma$, $\theta$ and $\lvert\sigma\rvert$, could lead to complementary constraints on whether or not a putative void is real or correctly identified.
These would only be partially independent from one another, since the four parameters are related to one another, with $\theta$ and $\lvert\sigma\rvert$ taking into account the evolution of the gravitational potential.
Deriving the weak-lensing parameters for an initial approximation, and then using the Sachs optical scalars for an analysis to higher accuracy could be one viable strategy.
Another extension would be to examine individual pairs of the best-matching {\thrD} and {\twD} voids from the current algorithm presented here to understand how their match could be improved; or alternatively, examine the worst-matching pairs to understand what obstructs the match and search for ways of avoiding the obstruction.

There are several advantages in detecting voids via their sky-plane effects.
Some of the fainter galaxies defining the walls of a void may be too faint to be detected in a given survey.
The tracing of dark matter by luminous matter is by a long chain of physical effects: baryonic matter has to be associated with the dark matter, and star formation history and feedback effects need to be modelled.
Geometric optics bypasses this causal chain, and should lead to inferences made with fewer assumptions.

\subsection{Projected void concentricity and obscuring cosmic web structures}

Projection of voids to be nearly concentric is expected in our simulation, since we integrate over the full box size of $L_{\mathrm{box}} = {\Lboxvalue} \mathrm{Mpc}/h$ and the largest intrinsic voids detected with the watershed algorithm have radii $R_\mathrm{eff} \sim 30 \mathrm{Mpc}/h$.
Our algorithm's only strategy that relates to the problem of projected void concentricity is to prefer larger to smaller radii (step \ref{algo-item-min-local-eta} in \SSS\ref{s-method-photo-voids}).
Figures \ref{fig-bestmatch-mass-def}, \ref{fig-bestmatch-weak-lens}, \ref{fig-bestmatch-expansion}, and \ref{fig-bestmatch-sigma} show that despite this, the {\twD} void radii tend to be lower than the intrinsic radii.
This empirical result would tend to favour keeping this step unchanged.

Our algorithm already has many parameters.
Extending it to allow successive multiple detections of walls could, in principle, lead to a higher rate of detecting the intrinsic voids.
Ideally, this should lead to a statistically significant correlation between the intrinsic and photometric void radii; in this work, our correlations in radii are insignificant (Table~\ref{t-match-significance}).
However, detecting multiple concentric walls would quite likely also lead to false detections.

Strategies for solving the problem of obscuring structures (in the absence of redshift information) are not obvious.
Gravitationally dense objects occupy little volume and still suffer from projection effects; voids dominate the volume and thus are strongly affected by projection effects.
A Bayesian approach as in \citet{Jeffrey2021DES3WeakLensing} would be worth exploring.

Since our simulation homogenises the foreground and background of the simulated volume, a real observational survey will include stronger levels of both projected void concentricity and obscuring cosmic web structures.

\section{Conclusion} \label{s-conclu}
In this work we have studied the two questions of whether voids in the cosmic web yield detectable information in projected variables, the surface overdensity $\Sigma$, the azimuthal averaged weak lensing shear $\bargamma$, the Sachs expansion $\theta$, and the Sachs shear $\lvert\sigma\rvert$, and vice versa, whether the sky-plane information can be used to infer the existence of the intrinsic {\thrD} voids.
We performed this using a cosmological $N$-body simulation starting from initial perturbations generated according to a standard initial power spectrum.
We carried out the analysis in a fully controlled software environment with full information about the dark matter distribution as well as the luminous matter distribution, which we modelled using galaxies built from a halo merger tree using semi-analytical tools.
We detected the intrinsic voids in the {\thrD} comoving spatial distribution of galaxies using a watershed void finder (\SSS\ref{s-method-3D-voids}).
The void detection in the projected plane (\SSS\ref{s-method-photo-voids}) is based on the assumption that the azimuthally averaged profiles of the four detector variables for the voids have shapes with predictable qualitative behaviour.
In the case of the surface overdensity $\Sigma$ and the two Sachs optical scalars  $\theta$ and $\lvert\sigma\rvert$, this expected shape is to start from a minimum at the centre of a void, gradually increase radially outwards, and increase sharply at the void's wall.
The weak-lensing shear $\bargamma$ is expected to start from zero, decrease, and increase to zero just past the void's wall.
Using a heuristically parametrised algorithm for detecting these profiles, adjusted individually for the four detector variables, we found positions and radii of {\twD} voids traced by these detectors.

We find roughly similar numbers of {\twD} voids traced by each of the four different detector variables, and in all cases, fewer voids than in the {\thrD} galaxy-traced distribution, as can be seen in Table~\ref{t-N-detections}.
There are two likely explanations.
First, when several intrinsic voids are nearly concentric in projection on the sky, our algorithm only detects one of these, since it is not designed to detect multiple walls.
Second, the foreground and background structures of the cosmic web, i.e. walls, filaments, clusters and other voids, obscure the signals associated with any single intrinsic void, making detection difficult.
The lower panels of Figs~\ref{fig-mass-def}, \ref{fig-weak-lens-gamma}, \ref{fig-theta}, and \ref{fig-optical-scalars-mag-sigma} show that the voids detected by us in the projected plane follow the assumed qualitative shapes well, giving confidence that our algorithm works as expected.
However, the same panels show that the corresponding mean profiles, using the centres and radial sizes of the {\thrD} intrinsic voids, but the detector variables in the projected plane, are weak.

We interpret these two effects -- the detection of fewer {\twD} voids than those known to exist in the {\thrD} spatial distribution, together with the weak mean profiles of the projected-plane detector variables centred at the intrinsic voids' locations -- as consistent with the undetected voids being (statistically) those that are either the most obscured or are concentric with the detected voids.

Given knowledge of the {\thrD} voids' centres, we find (Table~\ref{t-match-significance}, third column) that the detected {\twD} voids are signicantly closer than random to the {\thrD} voids' centres in the sky plane, for all four detector variables.
In other words, a survey with sufficient spectroscopic or photometric redshift information to detect voids should be usable to infer patterns of gravitational lensing through the voids that should be measurable using either weak-lensing shear or the Sachs optical scalars (answering question (ii) of \SSS\ref{s-method-matches} positively).

Conversely, if we only have a photometric survey that is blind, in the sense of having neither spectroscopic nor photometric redshift information, then we have established (Table~\ref{t-match-significance}, second column) that the {\twD} voids detected via weak-lensing tangential shear $\bargamma$, Sachs expansion $\theta$ or Sachs absolute shear $\lvert\sigma\rvert$ significantly reveal the true underlying {\thrD} void population (question (i) in \SSS\ref{s-method-matches}).
Use of the surface overdensity $\Sigma$ provides weaker evidence for revealing the sky-plane positions of the underlying void population.

While these results follow from significant correlations of voids' locations in the sky plane, we find no significant correlation for the radii.
The bottom panels of Figs~\ref{fig-bestmatch-mass-def}, \ref{fig-bestmatch-weak-lens}, \ref{fig-bestmatch-expansion}, and \ref{fig-bestmatch-sigma}, show that the {\twD} void radii tend to be lower than the intrinsic radii.
The lack of correlation and the generally lower radii are consistent with the problem of near concentric projection of multiple voids into the sky plane.

While our current results are exploratory, with several caveats as stated above, it does appear that gravitational lensing through individual voids should be observationally detectable.
Moreover, weak-lensing tangential shear and Sachs expansion and shear in future blind photometric surveys -- such as those provided by the Rubin C.\ Observatory's Legacy Survey of Space and Time (LSST; \citealt{SheldonBecker2023RubinObs}) -- should reveal the existence of intrinsic {\thrD} voids, yielding predictions that will be falsifiable by spectroscopic followup surveys such as those of the 4-metre Multi-Object Spectroscopy Telescope (4MOST; \citealt{deJong12VISTA4MOST,deJong19Messenger,Richard19CRS4MOST}) or the Dark Energy Spectroscopic Instrument (DESI; \citealt{Levi13DESI,Hahn2022DESIBG}).

\section*{Data availability statement}
The full reproducibility package for this paper, including input data and source software checksums, is available at {\projectzenodohref} and in live\footnote{\projectgitrepository} and archived\footnote{\projectgitrepositoryarchived} {\sc git} repositories.
Our main numerical results are available as DOI-identified records as indicated in the captions of Table~\ref{t-match-significance}, and Figs.~\ref{fig-bestmatch-mass-def}, \ref{fig-bestmatch-weak-lens}, \ref{fig-bestmatch-expansion}, and \ref{fig-bestmatch-sigma}.
The version of the source package used to produce this paper has the {\sc git} commit hash \projectversion.

\section*{Acknowledgments}
The authors wish to thank Mariana Jaber, Matteo Cinus and David Bacon for many useful comments and suggestions.
This work has been partly supported by the Polish MNiSW grant DIR/WK/2018/12.
This work has been partly supported by the Pozna\'n Supercomputing and Networking Center (PSNC) computational grant 537.
Part of this work was supported by Universitas Copernicana Thoruniensis in Futuro under NCBR grant POWR.03.05.00-00-Z302/17.
This work has been partly supported by the NCN grant 2022/45/N/ST9/01698.
This work was partly supported by the Australian Research Council through the Future Fellowship FT140101270.

{}We gratefully acknowledge the use of the following free-software programs and libraries:  1.23, Boost 1.77.0, Bzip2 1.0.8, convertctrees 0.0-522dac5, cosmdist 0.3.12, ctrees 1.01-7c20add, cURL 7.84.0, Dash 0.5.11-057cd65, diff-solver 0.0.1, Discoteq flock 0.4.0, Eigen 3.4.0, Expat 2.4.1, fftw2 2.1.5-4.2, FFTW 3.3.10 \citep{fftw}, File 5.42, Fontconfig 2.14.0, FreeType 2.11.0, Git 2.37.5, GNU Autoconf 2.71, GNU Automake 1.16.5, GNU AWK 5.1.1, GNU Bash 5.2-rc2, GNU Binutils 2.39, GNU Bison 3.8.2, GNU Compiler Collection (GCC) 12.1.0, GNU Coreutils 9.1, GNU Diffutils 3.8, GNU Findutils 4.9.0, GNU gettext 0.21, GNU gperf 3.1, GNU Grep 3.7, GNU Gzip 1.12, GNU Integer Set Library 0.24, GNU libiconv 1.17, GNU Libtool 2.4.7, GNU libunistring 1.0, GNU M4 1.4.19, GNU Make 4.3, GNU Multiple Precision Arithmetic Library 6.2.1, GNU Multiple Precision Complex library, GNU Multiple Precision Floating-Point Reliably 4.1.0, GNU Nano 6.4, GNU NCURSES 6.3, GNU Patch 2.7.6, GNU Readline 8.2-rc2, GNU Scientific Library 2.7, GNU Sed 4.8, GNU Tar 1.34, GNU Texinfo 6.8, GNU Wget 1.21.2, GNU Which 2.21, GPL Ghostscript 9.56.1, Help2man , ImageMagick 7.1.0-13, Less 590, Libffi 3.4.2, libICE 1.0.10, Libidn 1.38, Libjpeg 9e, Libpaper 1.1.28, Libpng 1.6.37, libpthread-stubs (Xorg) 0.4, libSM 1.2.3, Libtiff 4.4.0, libXau (Xorg) 1.0.9, libxcb (Xorg) 1.15, libXdmcp (Xorg) 1.1.3, libXext 1.3.4, Libxml2 2.9.12, libXt 1.2.1, LibYAML 0.2.5, Lzip 1.23, mpgrafic 0.3.19-4b78328, OpenBLAS 0.3.21, Open MPI 4.1.1, OpenSSL 3.0.5, PatchELF 0.13, Perl 5.36.0, pkg-config 0.29.2, podlators 4.14, Python 3.10.6, ramses-scalav 0.0-482f90f, revolver 0.0-3b15335, rockstar 0.99.9-RC3+-6d16969, sage 0.0-2be3027, util-Linux 2.38.1, util-macros (Xorg) 1.19.3, X11 library 1.8, XCB-proto (Xorg) 1.15, xorgproto 2022.1, xtrans (Xorg) 1.4.0, XZ Utils 5.4.1 and Zlib 1.2.11.
Part of this project used the following {\sc python} modules: Astropy 5.1 \citep{astropy2013,astropy2018},  BeautifulSoup 4.10.0,  Cycler 0.11.0, Cython 0.29.24 \citep{cython2011},  emcee 3.0.1,  Extension-Helpers 0.1,  HTML5lib 1.0.1,  Jinja2 3.0.3,  Kiwisolver 1.0.1,  MarkupSafe 2.0.1, Matplotlib 3.3.0 \citep{matplotlib2007}, Numpy 1.21.3 \citep{numpy2020},  Packaging 21.3,  Pillow 8.4.0,  pybind11 2.5.0,  PyERFA 2.0.0.1,  pyFFTW 0.12.0,  PyParsing 3.0.4,  python-dateutil 2.8.0,  python-installer 0.5.1,  PyYAML 6.0, Scipy 1.10.0 \citep{scipy2020},  Setuptools 58.3.0,  Setuptools-scm 3.3.3,  Six 1.16.0,  SoupSieve 1.8,  Webencodings 0.5.1 and  Wheel 0.37.0.
The typesetting of this paper was carried out using the following \LaTeX{} and related free-software packages: alegreya 64384 (revision), biber 2.19, biblatex 3.19, bitset 1.3, caption 62757 (revision), courier 61719 (revision), csquotes 5.2n, datetime 2.60, ec 1.0, enumitem 3.9, environ 0.3, etoolbox 2.5k, fancyhdr 4.1, fancyvrb 4.5a, fmtcount 3.07, fontaxes 1.0e, fontspec 2.8a, footmisc 6.0d, fp 2.1d, kastrup 15878 (revision), lastpage 2.0a, latexpand 1.7.2, letltxmacro 1.6, listings 1.9, logreq 1.0, mnras 3.1, mweights 53520 (revision), newtx 1.71, pdfcol 1.7, pdfescape 1.15, pdftexcmds 0.33, pgf 3.1.10, pgfplots 1.18.1, preprint 2011, setspace 6.7b, tcolorbox 6.0.1, tex 3.141592653, texgyre 2.501, times 61719 (revision), titlesec 2.14, trimspaces 1.1, txfonts 15878 (revision), ulem 53365 (revision), xcolor 2.14, xkeyval 2.9 and xstring 1.85.
{}

\end{document}